\documentclass[review,12pt]{elsarticle}
\biboptions{sort&compress}
\usepackage{amsmath,amssymb,bm}

\usepackage{amsthm}
\usepackage{graphicx}
\graphicspath{{figures/}}
\usepackage{booktabs,array}
\usepackage{placeins}
\usepackage{xcolor}
\usepackage{microtype}
\usepackage[colorlinks=true]{hyperref}
\usepackage{orcidlink}
\newcommand{\Wi}{\mathrm{Wi}}
\newcommand{\lmin}{\lambda_{\min}}
\newcommand{\khat}{\hat{\bm k}}
\newcommand{\A}{\bm{A}}
\newtheorem{proposition}{Proposition}
\newtheorem{corollary}{Corollary}
\theoremstyle{remark}
\newtheorem{remark}{Remark}
\begin{document}
\hypersetup{pdftitle={Loss of positive definiteness as constraint relaxation in high-Weissenberg-number computations},
pdfauthor={Yuan Yu; Lanjin Lian; Feiyang Chu}}
\begin{frontmatter}

\title{Loss of positive definiteness as constraint relaxation in high-Weissenberg-number computations}

\author[a,b,c]{Yuan Yu\corref{cor1}\,\orcidlink{0000-0001-5125-2492}}
\ead{yuyuan@xtu.edu.cn}
\author[a]{Lanjin Lian}
\author[a]{Feiyang Chu}
\cortext[cor1]{Corresponding author.}
\address[a]{School of Mathematics and Computational Science, Xiangtan University, Xiangtan 411105, China}
\address[b]{National Center for Applied Mathematics in Hunan, Xiangtan 411105, China}
\address[c]{Hunan Key Laboratory for Computation and Simulation in Science and Engineering, Xiangtan University, Xiangtan 411105, China}

\begin{abstract}
Loss of symmetric positive definiteness (SPD) in high-Weissenberg-number conformation calculations is commonly treated as a precursor to numerical breakdown. We show that it has a more direct meaning: the discrete update ceases to preserve a positivity constraint maintained by the continuum model, a departure we term constraint relaxation. For periodic Oldroyd--B flows with nonnegative stress diffusion, a classical solution starting from uniformly SPD data remains strictly SPD, whereas polynomial-closure discretizations can develop negative eigenvalues from equilibrium data. Whether this departure leads to breakdown depends on discrete stress--momentum coupling and on the extent to which a local growth threshold is exceeded, rather than on the eigenvalue sign itself. For a uniform frozen state with solvent viscosity ratio $0<\beta<1$, nondiffusive transverse eigenmodes grow only when $\lmin(\A)<-\beta/(1-\beta)$; the short-wave growth rate remains bounded, stress diffusion restricts growth to a finite wavenumber band, and two-dimensional constitutive relaxation contributes positively to the determinant balance wherever $\det\A<0$ and $\operatorname{tr}\A>0$. In lattice Boltzmann four-roll-mill simulations, we pair force-divergence and local stress-moment couplings with shifted-cone projection and a smooth softplus restoring source calibrated to $-\beta/(1-\beta)$. At $\beta=2/3$ ($\Wi=50$), time-consistent force-divergence coupling crosses the velocity stopping threshold even when the eigenvalues supplying polymer stress stay strictly positive or above $-1$ up to that crossing, whereas Guo force-divergence and local stress-moment schemes reach $t^*=120$ despite strongly negative eigenvalues. At $\beta=1/9$ under matched forcing, the unmodified stress-moment scheme terminates, whereas shifted-cone projection and smooth spectral restoring both maintain bounded unsteady continuation to $t^*=120$ without enforcing $\lmin(\A)>0$. Stabilization strategies should therefore control constraint relaxation together with discrete polymer-stress feedback, and numerical methods should be evaluated on both rather than on eliminating negative eigenvalues alone.
\end{abstract}

\begin{keyword}
high-Weissenberg-number problem \sep Oldroyd--B model \sep lattice Boltzmann method
\sep conformation tensor \sep positive definiteness \sep constraint relaxation
\end{keyword}
\end{frontmatter}

%==============================================================================
\section{Introduction}
\label{sec:intro}

In high-Weissenberg-number computations, a negative eigenvalue of the conformation tensor is commonly treated as a sign of impending numerical failure. In the Oldroyd--B model, $\A$ is a normalized second moment of the polymer configuration; a physically realizable conformation is therefore positive semidefinite, and a nondegenerate one is symmetric positive definite (SPD). A negative eigenvalue has one definite implication: the computed tensor no longer represents such a second moment. It does not, by itself, establish continuum ill-posedness or imminent termination of the discrete computation. Loss of physical realizability, local modal growth, and finite-time numerical termination are distinct claims. Under the periodic diffusive conditions studied here, a negative eigenvalue in a computation initialized with SPD data also has a precise numerical meaning: the discrete update has ceased to preserve the positive-definiteness constraint of the continuum model, a departure we term \emph{constraint relaxation}. Keeping these claims distinct determines both how SPD loss should be interpreted and what a numerical stabilization method should control.

The connection between loss of SPD and unbounded short-wave growth originates from theoretical results for specific model limits. Smooth, nondiffusive constitutive evolution preserves SPD initial data \cite{DM1986,Hulsen1990}, and nonnegativity persists when polymer-stress diffusion is added \cite{CK2012}. In the solvent-free, nondiffusive, inertial Maxwell limit ($\beta=0$), classical analyses related the elastic shear-wave speed and equation type to the conformation tensor \cite{DM1986,Rutkevich,Rutkevich1970,JRS1985}. An indefinite frozen state can then support short-wave perturbations whose growth rate becomes unbounded linearly with wavenumber \cite{JS1990}. Adding a Newtonian solvent changes the principal operator. For $\beta=\mu_s/(\mu_s+\mu_p)>0$, the momentum equation retains a parabolic part, and local well-posedness has been established for sufficiently smooth symmetric extra stress without assuming positive definiteness \cite{GS1990}. Joseph and Saut identified Newtonian viscosity as the natural regularization of the solvent-free short-wave instability \cite{JS1990,JS1986}. The Maxwell limit thus explains why many numerical methods enforce SPD, but it does not determine the dynamical consequences of a negative eigenvalue in an Oldroyd--B computation with solvent viscosity and stress diffusion.

Numerical studies also show that a local state constraint cannot by itself determine either stability or the reliability of every computed quantity. Cholesky, logarithmic, and square-root factorizations enforce positive \mbox{(semi-)}def\-i\-nite\-ness by construction \cite{VC2003,FK2004,FK2005,Balci2011}, while logarithmic free-energy and entropy formulations require strictly positive eigenvalues to remain well-defined \cite{HL2007}. Comparing four positive \mbox{(semi-)}def\-i\-nite reformulations across benchmark flows, Hulsen, Spanjaards, and Anderson \cite{HSA2021} nevertheless observed markedly different breakdown limits, interpreting them through Fattal and Kupferman's mechanism of steep, nearly exponential stress profiles in high-residence-time regions \cite{FK2004,FK2005}. Conversely, Capocci, Linkmann, and Morozov \cite{NoNeed} reported strongly negative conformation traces, directly indicating loss of SPD, in three-dimensional simplified Phan-Thien--Tanner simulations whose selected finite-time velocity and stress statistics remained close across grid resolutions. Beyond constraint enforcement and steep-profile resolution, two questions remain open for Oldroyd--B computations: how perturbations evolve locally about an indefinite state, and how such states interact with stress--momentum coupling in a discrete solver.

To answer these questions, we first contrast classical continuum evolution with discrete time stepping. Under periodic boundary conditions and with $\kappa\geq 0$, a classical Oldroyd--B solution starting from uniformly SPD data remains strictly inside the SPD cone throughout its interval of existence, whereas polynomial-closure discretizations initialized at $\A=\bm I$ can leave that cone through discrete \emph{constraint relaxation}, distinct from physical \emph{constitutive relaxation}. Extending the finite-solvent transverse dispersion baseline of Owens and Phillips \cite{OP2002} and Chen \cite{Chen2014} to include stress diffusion then shows that, about a prescribed uniform frozen state with $0<\beta<1$, nondiffusive transverse eigenmodes grow only when $\lmin(\A)<-\beta/(1-\beta)$, with positive solvent viscosity bounding the short-wave growth rate and stress diffusion restricting growth to a finite wavenumber band. A companion study analyzes static compliance and determines which wavevector orientation and transverse polarization destabilize first across Oldroyd--B, finitely extensible nonlinear elastic--Peterlin (FENE--P), and Giesekus closures \cite{YuRapids2026}. In two dimensions, constitutive relaxation drives $\det\A$ upward wherever $\det\A<0$ and $\operatorname{tr}\A>0$.

Once negative eigenvalues arise in steep stress layers, finite-horizon survival depends jointly on discrete polymer-stress feedback and on excursions beyond $-\beta/(1-\beta)$. In four-roll-mill computations with a dual-distribution two-relaxation-time lattice Boltzmann solver, we examine two algorithmic components. First, for discrete momentum--stress coupling, we contrast centered-difference force-divergence schemes \cite{Guo2002,ZXW2025} against derivative-free local second-moment Hermite stress injection \cite{Yu2026}; second, for threshold-anchored spectral control of polynomial closures, we compare Frobenius-orthogonal two-eigenvalue projection onto a shifted sub-threshold cone $\mathcal K_{-\gamma+\varepsilon}$ against a projection-free $C^\infty$ softplus spectral-restoring source $\bm S_{\mathrm{cone}}(\A)$. At $\Wi=50$ and $\beta=2/3$ (threshold $-2$), time-consistent force-divergence runs cross the velocity stopping threshold even when smooth spectral restoring is applied or when the conformation eigenvalues supplying polymer stress are held strictly positive or above $-1$ up to that crossing; by contrast, Guo force-divergence and local stress-moment runs reach $t^*=120$ despite $\lmin(\A)\approx -148$ to $-151$. At $\beta=1/9$ (threshold $-0.125$), matching the dimensionless driving force to the baseline terminates the unmodified stress-moment runs, whereas shifted-cone projection and the smooth spectral-restoring source keep all ten runs admitting $\lmin(\A)<0$, alongside both strict-SPD controls, bounded to $t^*=120$.

The remainder of the paper is organized as follows. Section~\ref{sec:theory} establishes periodic continuum SPD preservation, derives the finite-solvent frozen growth threshold, short-wave bound, and diffusive cutoff that motivate the shifted cone $\mathcal K_{-\gamma}$, and proves two-dimensional constitutive determinant recovery. Section~\ref{sec:methods} presents the viscoelastic lattice Boltzmann formulations, the five momentum-coupling schemes, the threshold-anchored projection and smooth spectral-restoring operators, and the four-roll-mill diagnostic protocol. Section~\ref{sec:verify} verifies the local dispersion predictions, transient growth peak timing on relaxing backgrounds, and smooth-flow operator consistency on steady Kolmogorov flow. Section~\ref{sec:causal} analyzes stress--momentum coupling at $\beta=2/3$, outside-cone field structure and determinant balance at $\Wi=20$, and finite-horizon stabilization at $\beta=1/9$. Section~\ref{sec:discussion} discusses regime limits, positive \mbox{(semi-)}def\-i\-nite reformulations, and implications for method evaluation, and Section~\ref{sec:concl} summarizes the conclusions.

%==============================================================================
\section{Continuum realizability, frozen-state growth threshold, and constitutive recovery}
\label{sec:theory}

\subsection{Continuum strict positivity and discrete constraint relaxation}
\label{sec:worlds}

We consider the incompressible diffusive Oldroyd--B equations \cite{Oldroyd1950} in conformation-tensor form on a periodic domain,
\begin{align}
\rho\,(\partial_t\bm u+\bm u\!\cdot\!\nabla\bm u)
&=-\nabla p+\mu_s\Delta\bm u+\nabla\!\cdot\!\bm\tau+\bm f,
\label{eq:mom}\\
\nabla\!\cdot\!\bm u&=0,\notag\\
\partial_t\A+\bm u\!\cdot\!\nabla\A
&=(\nabla\bm u)\A+\A(\nabla\bm u)^{\!\top}
-\frac{1}{\lambda}(\A-\bm I)+\kappa\Delta\A,\label{eq:conf}\\[-2pt]
&\hspace{3em}
\bm\tau=\frac{\mu_p}{\lambda}(\A-\bm I),\notag
\end{align}
where $\bm u$ is the velocity, $p$ is the pressure, $\A$ is the equilibrium-normalized conformation tensor, $\bm I$ is the identity tensor, $\bm\tau$ is the polymeric extra stress, and $(\nabla\bm u)_{ij}=\partial u_i/\partial x_j$. The material parameters are the fluid density $\rho$, Newtonian solvent viscosity $\mu_s$, polymer viscosity $\mu_p$, relaxation time $\lambda>0$, and stress diffusivity $\kappa\geq 0$, while $\bm f$ denotes the external body force and $\beta=\mu_s/(\mu_s+\mu_p)$ is the solvent viscosity ratio.

Microstructural realizability of $\A$ as a normalized polymer second moment defines the positive-semidefinite cone and its strictly positive-definite interior,
\begin{equation}
\mathcal K=\{\A=\A^{\!\top}:\A\succeq 0\},
\qquad
\mathcal K^{\circ}=\{\A=\A^{\!\top}:\A\succ 0\}.
\label{eq:realizability-cone}
\end{equation}
Throughout this paper, loss of SPD denotes strict departure $\lmin(\A)<0$ rather than boundary contact $\lmin(\A)=0$. Because the polynomial closure in Eqs.~\eqref{eq:mom}--\eqref{eq:conf} involves neither a matrix logarithm nor a square root, the PDE system remains algebraically well-defined on symmetric tensors outside $\mathcal K$.

When $\kappa=0$, preservation of $\mathcal K^{\circ}$ along smooth velocity fields is the Oldroyd--B specialization of Hulsen's material-trajectory theorem \cite{Hulsen1990}. With $\kappa>0$, Constantin and Kliegl \cite[Proposition~1 and the following remark]{CK2012} proved preservation of positive semidefiniteness (nonnegativity) for the diffusive Oldroyd--B conformation tensor on $\mathbb R^2$ via a parabolic minimum-eigenvalue argument and noted a higher-dimensional matrix extension. Because material-trajectory integration alone does not control $\kappa\Delta\A$, we give a self-contained periodic first-contact proof on $\mathbb T^d$ covering all $\kappa\geq 0$ under classical regularity.

\begin{proposition}[Continuum realizability]
\label{prop:continuum-cone}
Let $\Omega=\mathbb T^d$, $\lambda>0$, and $\kappa\geq 0$. Suppose that $\bm u$ and the symmetric tensor $\A$ form a classical solution of Eq.~\eqref{eq:conf} on $[0,T]$ and that $\A(\bm x,0)$ is uniformly positive definite. Then
\begin{equation*}
\A(\bm x,t)\in\mathcal K^{\circ}
\qquad\text{for every }\bm x\in\Omega\text{ and }0\leq t\leq T.
\end{equation*}
\end{proposition}

\begin{proof}
Suppose to the contrary that $t_*\in(0,T]$ is the first time at which $\A$ touches $\partial\mathcal K$. Then there exist $\bm x_*\in\Omega$ and a fixed unit vector $\bm q$ such that $\A(\bm x_*,t_*)\bm q=\bm 0$. Define the scalar projection $\varphi(\bm x,t)=\bm q^{\!\top}\A(\bm x,t)\bm q$. On the boundaryless periodic torus $\mathbb T^d$, first contact implies that $\varphi(\cdot,t_*)$ attains its global minimum $0$ at $\bm x_*$, giving $\nabla\varphi(\bm x_*,t_*)=\bm 0$ (and $\Delta\varphi(\bm x_*,t_*)\geq 0$ when $\kappa>0$). Projecting Eq.~\eqref{eq:conf} onto $\bm q$ annihilates both upper-convected stretching terms via $\A(\bm x_*,t_*)\bm q=\bm 0$, leaving
\begin{equation}
\partial_t\varphi(\bm x_*,t_*)
=-\bm u\!\cdot\!\nabla\varphi+\frac{1}{\lambda}
+\kappa\Delta\varphi
\geq\frac{1}{\lambda}>0,
\label{eq:first-contact}
\end{equation}
which contradicts first arrival at $0$ from positive values.
\end{proof}

For a spatial discretization on $N_h$ nodes, write the discrete realizability cone as
\begin{equation}
\mathcal K_h=\prod_{j=1}^{N_h}\mathcal K,
\label{eq:discrete-cone}
\end{equation}
let $Z_h^{n+1}=\Phi_h(Z_h^n)$ denote the one-step coupled discrete update, and let $\A_h^n=\mathcal P_A Z_h^n$ extract its nodal conformation field. Universal discrete cone invariance requires that every admissible state $Z_h$ with $\mathcal P_A Z_h\in\mathcal K_h$ satisfy
\begin{equation}
\mathcal P_A\Phi_h(Z_h)\in\mathcal K_h.
\label{eq:discrete-invariance}
\end{equation}
From initial state $Z_h^0$, the trajectory visits the reachable conformation set
\begin{equation}
\mathcal R_h^A(Z_h^0)
=\{\mathcal P_A\Phi_h^n(Z_h^0):n=0,1,2,\ldots\}.
\label{eq:reachable-set}
\end{equation}
A coupled discretization exhibits \emph{constraint relaxation} whenever an equilibrium initialization exits $\mathcal K_h$,
\begin{equation}
\A_{h,j}^0=\bm I\quad\text{for every }j,
\qquad
\A_h^0\in\mathcal K_h,
\qquad
\exists\,n,j:\ \lmin(\A_{h,j}^n)<0,
\label{eq:discrete-relaxation-test}
\end{equation}
equivalently $\mathcal R_h^A(Z_h^0)\cap\mathcal K_h^c\neq\varnothing$. One trajectory satisfying Eq.~\eqref{eq:discrete-relaxation-test} suffices to disprove universal discrete cone invariance.

\begin{remark}
\label{rem:terminology}
We reserve \emph{constraint relaxation} exclusively for this discrete departure from $\mathcal K_h$, keeping \emph{constitutive relaxation} for $-(\A-\bm I)/\lambda$, \emph{two-relaxation-time} for the lattice Boltzmann collision operator, and \emph{smooth spectral-restoring source} for $\bm S_{\mathrm{cone}}(\A)$ in Section~\ref{sec:operators}. Section~\ref{sec:first-cross} locates the sub-step stage at which the discrete state first leaves $\mathcal K_h$.
\end{remark}

\subsection{Frozen-state transverse dispersion, diffusive cutoff, and shifted cone}

Once discrete constraint relaxation produces an indefinite state, its local linear response can be isolated around a homogeneous background. Although classical continuum trajectories remain in $\mathcal K^{\circ}$, linearizing the polynomial-closure PDE at a prescribed symmetric tensor $\A_0\notin\mathcal K$ directly characterizes the local algebraic feedback acting on discrete states once they leave $\mathcal K_h$. In an unforced periodic box with $\bm u=\bm 0$, a spatially uniform conformation state $\overline{\A}(0)\neq\bm I$ relaxes homogeneously according to
\begin{equation}
\overline{\A}(t)=\bm I+
\big(\overline{\A}(0)-\bm I\big)e^{-t/\lambda}.
\label{eq:relaxing-background}
\end{equation}
Freezing $\A_0=\overline{\A}(t_0)$ at a selected instant $t_0$ evaluates the instantaneous linear response of plane-wave perturbations $(\delta\bm u,\delta\!\A,\delta p)\propto\exp(i\bm k\cdot\bm x+\sigma t)$ with wavevector $\bm k\neq\bm 0$, magnitude $k=|\bm k|>0$, unit direction $\khat=\bm k/k$, and temporal growth rate $\operatorname{Re}\sigma$. Because $\bm k\cdot\delta\bm u=0$, velocity perturbations are purely transverse to $\bm k$. Eliminating the mixed transverse--longitudinal conformation modes that couple to transverse momentum (Appendix~\ref{app:disp}) yields, for each transverse polarization,
\begin{equation}
(\rho\sigma+\mu_s k^2)
\Big(\sigma+\frac{1}{\lambda}+\kappa k^2\Big)
=\frac{\mu_p}{\lambda}\,k^2\,
\big(-\khat^{\!\top}\!\A_0\khat\big).
\label{eq:disp}
\end{equation}
At $\kappa=0$, Eq.~\eqref{eq:disp} recovers the Oldroyd--B quadratic dispersion relation in Owens and Phillips \cite[subsection~3.3.1, Eqs.~(3.37)--(3.39), pp.~57--58]{OP2002} upon identifying the shifted stress $(\mu_p/\lambda)\A_0$. Owens and Phillips used this relation to show that the Hadamard blowup of the solvent-free limit does not occur when $\mu_s>0$, and Chen \cite[Appendix~B, Eqs.~(B.9)--(B.10), p.~156]{Chen2014} derived the two-dimensional directional threshold and stable sub-threshold negative range (see also \cite{YuRapids2026}). Taking this nondiffusive threshold as the analytical baseline, Proposition~\ref{prop:threshold} collects the arbitrary-dimension criterion $-\mu_s/\mu_p$, stable indefinite strip $[-\mu_s/\mu_p,0)$, and short-wave plateau in self-contained form, and Eq.~\eqref{eq:disp} extends the relation to $\kappa\geq 0$.

\begin{proposition}[Growth threshold and short-wave bound, $\mu_s>0$]
\label{prop:threshold}
Let $\mu_s,\mu_p>0$ (equivalently $0<\beta<1$) and $\kappa=0$. For any $k>0$, a fixed wavevector direction $\khat$ admits a growing root if and only if
\begin{equation}
\khat^{\!\top}\!\A_0\khat< -\frac{\mu_s}{\mu_p}.
\label{eq:threshold-directional}
\end{equation}
When Eq.~\eqref{eq:threshold-directional} holds with $\rho>0$, the unstable root increases monotonically with $k$ toward the finite high-wavenumber plateau
\begin{equation}
\sigma_\infty(\khat)
=\frac{1}{\lambda}
\left[
\frac{\mu_p}{\mu_s}
\big(-\khat^{\!\top}\!\A_0\khat\big)-1
\right].
\label{eq:rate-directional}
\end{equation}
Minimizing over $|\khat|=1$ shows that at least one wavevector direction grows if and only if
\begin{equation}
\lmin(\A_0)<-\frac{\mu_s}{\mu_p}
=-\frac{\beta}{1-\beta},
\label{eq:threshold}
\end{equation}
in which case the maximum short-wave growth rate is
\begin{equation}
\sigma_{\max}
=\frac{1}{\lambda}
\left[\frac{\mu_p}{\mu_s}\,|\lmin(\A_0)|-1\right].
\label{eq:rate}
\end{equation}
The instability threshold in Eqs.~\eqref{eq:threshold-directional} and~\eqref{eq:threshold} is independent of fluid inertia.
\end{proposition}

\begin{proof}
Writing Eq.~\eqref{eq:disp} as $\rho\sigma^2+b\sigma+c=0$ gives $b>0$, so a root with $\operatorname{Re}\sigma>0$ exists if and only if $c<0$, which at $\kappa=0$ reduces to Eq.~\eqref{eq:threshold-directional}. Setting $K=\mu_p(-\khat^{\!\top}\!\A_0\khat)/\lambda$ and $q=k^2$ for $\rho>0$ inverts Eq.~\eqref{eq:disp} along the growing branch as
\begin{equation}
q(\sigma)=
\frac{\rho\sigma(\sigma+1/\lambda)}
{K-\mu_s(\sigma+1/\lambda)},
\label{eq:q-of-sigma}
\end{equation}
which maps $(0,\sigma_\infty)$ monotonically onto $(0,\infty)$, bounding $\sigma$ by $\sigma_\infty(\khat)$ (attained identically for every $k>0$ when $\rho=0$).
\end{proof}

Newtonian solvent viscosity thus converts unbounded Maxwell short-wave blowup into a thresholded, rate-bounded response rather than stabilizing all indefinite states.

\begin{corollary}[Solvent-free, nondiffusive limit]
\label{cor:ucm}
If $\mu_s=0$, $\kappa=0$, $\rho>0$, and $\khat^{\!\top}\!\A_0\khat<0$, Eq.~\eqref{eq:disp} reduces to
\begin{equation*}
\rho\sigma\left(\sigma+\frac{1}{\lambda}\right)
=\frac{\mu_p}{\lambda}k^2
|\khat^{\!\top}\!\A_0\khat|,
\qquad
\sigma\sim k
\sqrt{\frac{\mu_p}{\rho\lambda}
|\khat^{\!\top}\!\A_0\khat|}.
\end{equation*}
Linearly divergent short-wave growth occurs only in this inertial, solvent-free, nondiffusive limit; retaining $\kappa>0$ at $\mu_s=0$ caps the high-$k$ rate at $\mu_p|\khat^{\!\top}\!\A_0\khat|/(\rho\kappa\lambda)$.
\end{corollary}

With both $\mu_s>0$ and $\kappa>0$, the condition $c<0$ in Eq.~\eqref{eq:disp} becomes
\begin{equation}
\mu_s(1+\kappa\lambda k^2)
<\mu_p\big(-\khat^{\!\top}\!\A_0\khat\big),
\label{eq:diffusive-threshold}
\end{equation}
confining growth along any supercritical direction $\khat$ to wavenumbers smaller than the diffusive cutoff
\begin{equation}
k^2<k_c^2=
\frac{(\mu_p/\mu_s)
|\khat^{\!\top}\!\A_0\khat|-1}{\kappa\lambda}.
\label{eq:cutoff}
\end{equation}
Because stress diffusion alters the governing constitutive balance, this wavenumber cutoff reflects a modified physical model rather than a purely numerical filter \cite{SB1995}.

Relative to the nondiffusive frozen-state threshold $-\beta/(1-\beta)$, we quantify pointwise threshold exceedance by
\begin{equation}
\chi(\A)=\frac{\mu_p}{\mu_s}
\big|\min(\lmin(\A),0)\big|-1,
\label{eq:Xi}
\end{equation}
so $\chi(\A)>0$ marks supercritical states with nondiffusive plateau $\sigma_{\max}=\chi(\A)/\lambda$.

\begin{remark}
\label{rem:lattice-nonnormal}
On a finite periodic box $[0,L)^d$ with fundamental wavenumber $k_{\min}=2\pi/L$, wavevectors are restricted to the discrete reciprocal lattice, so modal growth requires at least one admissible nonzero lattice vector $\bm k$ satisfying
\[
\khat^{\!\top}\A_0\khat<
-\frac{\beta}{1-\beta}(1+\kappa\lambda|\bm k|^2).
\]
Exceeding the continuous-wavenumber bound $k_c>k_{\min}$ is therefore necessary but not sufficient unless a reciprocal-lattice vector falls inside the supercritical sector. Moreover, $\chi(\A)$ characterizes frozen normal-mode growth rather than nonnormal transient amplification in spatially inhomogeneous flows.
\end{remark}

To interpolate systematically between strict SPD preservation and the nondiffusive frozen-state threshold $-\beta/(1-\beta)$ for $0<\beta<1$, we define for $c_\gamma\in[0,1)$ the affine \emph{shifted cone}
\begin{equation}
\mathcal K_{-\gamma}=\mathcal K-\gamma\bm I=\{\A=\A^{\!\top}:\A+\gamma\bm I\succeq 0\},
\qquad
\gamma=c_\gamma\frac{\beta}{1-\beta}=c_\gamma\frac{\mu_s}{\mu_p},
\label{eq:shifted-cone}
\end{equation}
where $\mathcal K_0=\mathcal K$ has interior $\mathcal K^{\circ}$, and $\mathcal K_{-\gamma}^{\circ}$ denotes the open shifted cone with $\lmin(\A)>-\gamma$. Every tensor $\A\in\mathcal K_{-\gamma}$ satisfies $\chi(\A)\leq c_\gamma-1<0$ on the nondiffusive frozen branch while admitting negative eigenvalues in $[-\gamma,0)$ whenever $c_\gamma>0$.

\subsection{Constitutive recovery in the two-dimensional determinant balance}
\label{sec:selfheal}

Whereas transverse dispersion governs whether perturbations grow about a frozen indefinite tensor, the homogeneous constitutive relaxation term acts directly on the tensor invariants themselves. For smooth nondiffusive three-dimensional flows, Hulsen \cite[Eq.~(18)]{Hulsen1990} derived the invariant evolution equation for $\det\A$ to establish preservation of $\mathcal K^{\circ}$. To evaluate how Oldroyd--B constitutive relaxation acts outside $\mathcal K$ in two dimensions without invoking $\A^{-1}$ or $\log\A$, we use the polynomial adjugate identities $\mathrm d\det\A=\operatorname{tr}(\operatorname{adj}\A\,\mathrm d\A)$ and $\operatorname{tr}(\operatorname{adj}\A)=\operatorname{tr}\A$ ($d=2$), which hold across singular and indefinite states; applying them to Eq.~\eqref{eq:conf} with stress diffusion retained yields
\begin{equation}
\frac{\mathrm D\det\A}{\mathrm Dt}
=2(\nabla\!\cdot\!\bm u)\det\A
-\frac{2}{\lambda}\det\A
+\frac{\operatorname{tr}\A}{\lambda}
+\kappa\operatorname{tr}
\left(\operatorname{adj}\A\,\Delta\A\right)
\qquad\text{(2D)}.
\label{eq:det2d}
\end{equation}
Appendix~\ref{app:det} derives Eq.~\eqref{eq:det2d}, contrasts it with its three-dimensional counterpart, and constructs an explicit three-dimensional indefinite tensor with $\operatorname{tr}\A>0$ whose relaxation derivative of $\det\A$ is negative.

For $\nabla\!\cdot\!\bm u=0$ in two dimensions, the constitutive-relaxation term in Eq.~\eqref{eq:det2d} is
\begin{equation}
H_{\mathrm{rel}}
=\frac{\operatorname{tr}\A-2\det\A}{\lambda},
\label{eq:relaxation-contribution}
\end{equation}
which satisfies $H_{\mathrm{rel}}>0$ whenever $\det\A<0$ and $\operatorname{tr}\A>0$ (and at regular boundary points where $\det\A=0<\operatorname{tr}\A$). Within this quadrant, constitutive relaxation always pushes $\det\A$ upward, whereas full re-entry into $\mathcal K$ depends on competition with advection, stretching, and diffusion. In the absence of transport, stretching, and diffusion, Eq.~\eqref{eq:relaxing-background} implies the dimension-independent eigenvalue relaxation law
\begin{equation}
\lambda_i(t)=1+\big(\lambda_i(0)-1\big)e^{-t/\lambda},
\label{eq:pure-relaxation}
\end{equation}
which drives any initial eigenvalue $\lmin(\A_0)<0$ monotonically across zero at $t_{\mathrm{cross}}=\lambda\ln(1-\lmin(\A_0))$. Once a field already contains outside-cone points at $\kappa>0$, $\kappa\operatorname{tr}(\operatorname{adj}\A\,\Delta\A)$ in Eq.~\eqref{eq:det2d} has no fixed pointwise sign; this is consistent with Proposition~\ref{prop:continuum-cone}, which uses $\Delta\varphi\geq 0$ only at a spatial minimum upon first boundary contact from inside $\mathcal K^{\circ}$.

%==============================================================================
\section{Viscoelastic lattice Boltzmann formulations and threshold-anchored spectral operators}
\label{sec:methods}

\subsection{Nondimensionalization and dual-distribution solvers}

On the periodic square $\Omega=[0,L_b)^2$ with fundamental wavenumber $k_b=2\pi/L_b$, we scale length by $\ell_c=k_b^{-1}$, velocity by $U_c$, and time by $T_c=\ell_c/U_c$, reporting dimensionless times $t^*=t/T_c$ and modal growth rates $\sigma T_c$. The governing dimensionless parameters define the Reynolds number $\mathrm{Re}$, Weissenberg number $\Wi$, solvent viscosity ratio $\beta$, Schmidt number $\mathrm{Sc}$ (for $\kappa>0$), and dimensionless stress diffusivity $\kappa^*$,
\begin{equation}
\begin{aligned}
\mathrm{Re}&=\frac{\rho U_c\ell_c}{\mu_s+\mu_p}, &
\Wi&=\frac{\lambda}{T_c}, &
\beta&=\frac{\mu_s}{\mu_s+\mu_p},\\
\mathrm{Sc}&=\frac{\nu_s}{\kappa}, &
\kappa^*&=\frac{\kappa}{U_c\ell_c}=\frac{\beta}{\mathrm{Re}\,\mathrm{Sc}}, & &
\end{aligned}
\label{eq:nondim-groups}
\end{equation}
where $\nu_s=\mu_s/\rho$ is the kinematic solvent viscosity. Equation~\eqref{eq:nondim-groups} shows that lowering $\beta$ at fixed $\mathrm{Re}$ and $\mathrm{Sc}$ proportionally reduces $\kappa^*$, a parameter coupling separated explicitly in the low-solvent controls of Section~\ref{sec:lowbeta}. In the weakly compressible lattice Boltzmann solver, both the hydrodynamic and conformation lattices share the time step $\Delta t$, unit lattice speed $\Delta x/\Delta t=1$, and D2Q9 lattice sound speed $c_s=1/\sqrt{3}$; fixing the reference Mach number at $\mathrm{Ma}=U_c/c_s=10^{-2}$ keeps the local Mach number $|\bm u|/c_s=10^{-2}|\bm u|/U_c$ small during normal operation and marks loss of low-Mach validity when velocities grow large.

The primary coupled solver advances one D2Q9 hydrodynamic distribution $f_i$ for $(\rho,\bm u)$ alongside three D2Q9 advection--diffusion distributions $g_{ai}$ ($a\in\{xx,xy,yy\}$) for the independent components of the symmetric conformation tensor $\A$, with dimensionless relaxation times $\tau_s=\tfrac{1}{2}+\nu_s/(c_s^2\Delta t)$ and $\tau_p=\tfrac{1}{2}+\kappa/(c_s^2\Delta t)$. The hydrodynamic equation employs a Hermite-regularized two-relaxation-time (TRT) collision operator \cite{Ginzburg2008} with magic parameter $\Lambda_s=1/4$ (retaining non-equilibrium Hermite orders $1,2$ at rate $\Omega_{s1}=1/\tau_s$ and order $3$ at $\Omega_{s2}$, with $\Omega_{s4}=1$), a third-order Hermite equilibrium \cite{SYC2006}, and a cubic Galilean-invariance correction using centered-difference $\partial_\alpha(\rho u_\alpha^3)$, while each conformation component uses a second-order Hermite equilibrium, Hermite-regularized TRT collision with $\Lambda_p=10^{-6}$ (retaining order $1$ at $\Omega_{p1}=1/\tau_p$ and order $2$ at $\Omega_{p2}$, with $\Omega_{p3}=\Omega_{p4}=1$), and centered finite differences for $\nabla\bm u$ in the constitutive stretching source (sub-step ordering and mesoscopic source terms in Appendix~\ref{app:intervention}). Macroscopic component reconstruction $A_a=\sum_i g_{ai}$ assembles $\A$ and supplies the discrete polymeric extra stress
\begin{equation}
\bm\sigma_p\equiv\bm\tau=\frac{\mu_p}{\lambda}(\A-\bm I),
\label{eq:discrete-tau}
\end{equation}
where $\bm\tau$ is the continuum stress in Eq.~\eqref{eq:mom}; this polynomial relation remains well-defined when $\A\notin\mathcal K$, enabling direct evaluation of stress feedback in Section~\ref{sec:coupling}.

To measure local modal growth rates independently without weakly compressible lattice effects, we pair this solver with an incompressible Fourier pseudospectral code on $[0,2\pi)^2$ using $2/3$-rule dealiasing and fourth-order Runge--Kutta integration, with local-test initialization and fitting windows listed in Appendix~\ref{app:methods} (Table~\ref{tab:local-protocol}).

\subsection{Momentum--stress coupling discretizations: force divergence versus local stress-moment injection}
\label{sec:coupling}

At matched physical parameters, grid spacing, and conformation update, we compare five hydrodynamic coupling schemes that systematically vary how $\bm\sigma_p$ and the background body force $\bm F_b$ enter the momentum lattice equation. Let $w_i$ and $\bm c_i$ denote the D2Q9 weights and discrete velocities, and let $\Omega_s=1/\tau_s$ denote the symmetric hydrodynamic relaxation rate with dimensionless relaxation time $\tau_s$. Force-based lattice Boltzmann updates build upon the second-order velocity-space forcing rate $\mathcal G_i(\bm F,\bm u)$ and, in time-consistent formulations \cite{ZXW2025}, the per-step quadratic population correction $R_i(\bm F)$, given by
\begin{align}
\mathcal G_i(\bm F,\bm u)
&=w_i\left[
\frac{(\bm c_i-\bm u)\mathbin{\cdot}\bm F}{c_s^2}
+\frac{(\bm u\mathbin{\cdot}\bm c_i)(\bm c_i\mathbin{\cdot}\bm F)}{c_s^4}
\right],\nonumber\\
R_i(\bm F)
&=\frac{\Delta t^2w_i}{8}\left[
\frac{(\bm c_i\mathbin{\cdot}\bm F)^2}{\rho c_s^4}
-\frac{|\bm F|^2}{\rho c_s^2}
\right],
\label{eq:forcing-kernels}
\end{align}
where $\bm F$ is the vector force supplied to the forcing operator and $R_i(\bm F)$ enters both the non-equilibrium shift $f_i^{\mathrm{neq}}=f_i-f_i^{\mathrm{eq}}-R_i(\bm F)$ and the post-collision population increment. Whenever polymer stress is lumped into $\bm F$, $R_i(\bm F)$ introduces pointwise quadratic products of $\bm F$ into the momentum update.

In the two \emph{force-divergence} (FD) schemes, the polymer force $\bm F_p=\nabla\!\cdot\!\bm\sigma_p$ is first differentiated spatially by second-order centered differences and added to $\bm F_b$ as $\bm F=\bm F_b+\bm F_p$:
\begin{enumerate}
\item \textbf{FD-TC} (time-consistent force divergence \cite{ZXW2025}) evaluates the unshifted velocity $\bm u=(\sum_i\bm c_i f_i)/\rho$ and applies the post-collision increment $\Delta t\,\mathcal G_i(\bm F_b+\bm F_p,\bm u)+R_i(\bm F_b+\bm F_p)$ with no direct stress-moment source.
\item \textbf{FD-Guo} (Guo force divergence \cite{Guo2002}) uses the half-force shifted velocity $\bm u=(\sum_i\bm c_i f_i+\tfrac{1}{2}\Delta t\,\bm F)/\rho$ and applies $\Delta t(1-\Omega_s/2)\mathcal G_i(\bm F_b+\bm F_p,\bm u)$, omitting $R_i$.
\end{enumerate}

In the three \emph{stress-moment} (SM) schemes, the momentum update uses no spatial derivative of $\bm\sigma_p$. Instead, using the second-order Hermite tensor $\bm H_i^{(2)}=\bm c_i\bm c_i-c_s^2\bm I$, nodal polymer stress enters the hydrodynamic collision update directly at rate \cite{Yu2026}
\begin{equation}
T_i=-\frac{w_i}{2c_s^4\tau_s\Delta t}\left(
H^{(2)}_{i,xx}\sigma_{p,xx}
+2H^{(2)}_{i,xy}\sigma_{p,xy}
+H^{(2)}_{i,yy}\sigma_{p,yy}\right),
\label{eq:Ti}
\end{equation}
where $xx,xy,yy$ index Cartesian tensor components. Because the second velocity moment of $\Delta t\,T_i$ equals $-\bm\sigma_p/\tau_s$ locally on the lattice, subsequent lattice streaming generates $\nabla\!\cdot\!\bm\sigma_p$ at the macroscopic level without a separate finite-difference stencil. Across all three SM variants, the body-force kernel receives only $\bm F_b$:
\begin{enumerate}
\setcounter{enumi}{2}
\item \textbf{SM-1} (baseline stress-moment scheme \cite{Yu2026}) uses the half-force shift with $\bm F_b$ and applies $\Delta t(1-\Omega_s/2)(w_i/c_s^2)(\bm c_i\mathbin{\cdot}\bm F_b)+\Delta t\,T_i$.
\item \textbf{SM-Guo} uses the half-force shift with $\bm F_b$ and applies $\Delta t(1-\Omega_s/2)\mathcal G_i(\bm F_b,\bm u)+\Delta t\,T_i$.
\item \textbf{SM-TC} uses the unshifted velocity $\bm u=(\sum_i\bm c_i f_i)/\rho$ and applies $\Delta t\,\mathcal G_i(\bm F_b,\bm u)+R_i(\bm F_b)+\Delta t\,T_i$.
\end{enumerate}
Because $T_i$ is identical across SM-1, SM-Guo, and SM-TC, comparing these three schemes isolates the discretization of $\bm F_b$ under identical local stress-moment feedback, whereas contrasting FD-TC with SM-TC replaces centered-difference $\nabla\!\cdot\!\bm\sigma_p$ with $T_i$ while removing $\bm F_p$ from $R_i$ in Eq.~\eqref{eq:forcing-kernels}. Appendix~\ref{app:intervention} specifies two auxiliary FD-TC sensitivity controls: an order-$8$ spatial conformation filter and a polymer-force gain factor $g\bm F_p$.

\subsection{Threshold-anchored shifted-cone projection and smooth spectral-restoring source}
\label{sec:operators}

Separating strict positive definiteness from confinement relative to the nondiffusive frozen-state threshold $-\beta/(1-\beta)$ motivates two spectral operators anchored to the shifted cone $\mathcal K_{-\gamma}$ of Eq.~\eqref{eq:shifted-cone}, with $\gamma=c_\gamma\beta/(1-\beta)$ and $c_\gamma\in[0,1)$. Write the orthogonal eigendecomposition of the reconstructed symmetric nodal tensor $\A$ ($A_a=\sum_i g_{ai}$) as $\A=a_1\bm q_1\bm q_1^{\!\top}+a_2\bm q_2\bm q_2^{\!\top}$ with ordered eigenvalues $a_1\leq a_2$ and orthonormal eigenvectors $\{\bm q_1,\bm q_2\}$.

The first operator is the \emph{shifted two-eigenvalue floor}, which projects $\A$ orthogonally in the Frobenius norm onto $\mathcal K_{-\gamma+\varepsilon}$ prior to polymer-stress evaluation, defined by
\begin{equation}
\A^{\mathrm{new}}
=\Pi_{-\gamma+\varepsilon}(\A)
=\sum_{k=1}^2 \max(a_k,\ell_{\mathrm{floor}})\,\bm q_k\bm q_k^{\!\top},
\qquad
\ell_{\mathrm{floor}}=-\gamma+\varepsilon,
\label{eq:shifted-proj}
\end{equation}
with $\varepsilon=10^{-8}$. Setting $c_\gamma=0$ ($\ell_{\mathrm{floor}}=\varepsilon$) recovers the strict-SPD two-eigenvalue floor, whereas $c_\gamma\in(0,1)$ with $\gamma>\varepsilon$ permits negative eigenvalues down to $-\gamma+\varepsilon$ while keeping every nodal eigenvalue greater than $-\beta/(1-\beta)$. A one-sided floor, examined as an auxiliary control in Section~\ref{sec:clipping} and Appendix~\ref{app:intervention}, bounds only $a_1$ and leaves $\A$ indefinite whenever $a_1\leq a_2<0$. By contrast, Eq.~\eqref{eq:shifted-proj} acts on both eigenvalues: it is $C^0$ continuous and nonexpansive on symmetric $2\times 2$ matrices and guarantees $\A^{\mathrm{new}}\in\mathcal K_{-\gamma+\varepsilon}$. Each component increment $\Delta A_a=A_a^{\mathrm{new}}-A_a$ is then written back isotropically to the conformation populations before the next step as
\begin{equation}
g_{ai}\leftarrow g_{ai}+w_i\Delta A_a.
\label{eq:pop-writeback}
\end{equation}
Because $\sum_i w_i=1$, $\sum_i\bm c_i w_i=\bm 0$, and $\sum_i\bm c_i\bm c_i w_i=c_s^2\bm I$, Eq.~\eqref{eq:pop-writeback} shifts the zeroth moment by $\Delta A_a$ and the raw diagonal second moment by $c_s^2\Delta A_a\bm I$, while leaving the raw first velocity moment $\sum_i\bm c_i g_{ai}$ and the full second-order Hermite moment $\sum_i\bm H_i^{(2)}g_{ai}$ (including off-diagonal raw second moments) unchanged (shifting the non-equilibrium first moment relative to the updated equilibrium $g_{ai}^{\mathrm{eq}}$ by $-\bm u\Delta A_a$).

The second operator eliminates non-smooth clipping altogether by omitting Eq.~\eqref{eq:shifted-proj} and augmenting the constitutive source during conformation collision with an isotropic L\"owner \emph{smooth spectral-restoring source}, given by
\begin{equation}
\bm S_{\mathrm{cone}}(\A)
=\sum_{k=1}^2 g_{\mathrm{cone}}(a_k)\,\bm q_k\bm q_k^{\!\top},
\qquad
g_{\mathrm{cone}}(a)
=\frac{\delta_{\mathrm{cone}}}{\tau_{\mathrm{cone}}}
\ln\!\left[1+\exp\!\left(\frac{-\gamma-a}{\delta_{\mathrm{cone}}}\right)\right],
\label{eq:Scone}
\end{equation}
where $\delta_{\mathrm{cone}}=0.05\max(\gamma,0.05)$ sets the softplus transition width and $\tau_{\mathrm{cone}}=\max(r_{\mathrm{cone}}\min(\lambda,T_c),\,2.5\Delta t)$ sets the restoring timescale for dimensionless ratio $r_{\mathrm{cone}}\in\{0.01,0.05\}$. Because the idealized softplus kernel $g_{\mathrm{cone}}(a)$ in Eq.~\eqref{eq:Scone} is real analytic, $\bm S_{\mathrm{cone}}(\A)$ is $C^\infty$ on symmetric $2\times 2$ tensors, has Fr\'echet derivative $D\bm S_{\mathrm{cone}}(a\bm I)[\bm H]=g_{\mathrm{cone}}'(a)\bm H$ at a repeated eigenvalue $a_1=a_2=a$ without eigenvector singularity, and decays exponentially inside $\mathcal K_{-\gamma}^{\circ}$ once $a_1+\gamma\gg\delta_{\mathrm{cone}}$.

\begin{remark}
\label{rem:modified-pde}
Adding $\bm S_{\mathrm{cone}}(\A)$ modifies the macroscopic constitutive equation to $\partial_t\A+\bm u\!\cdot\!\nabla\A-(\nabla\bm u)\A-\A(\nabla\bm u)^{\!\top}=-(\A-\bm I)/\lambda+\kappa\Delta\A+\bm S_{\mathrm{cone}}(\A)$, acting primarily outside $\mathcal K_{-\gamma}$ for smooth $C^\infty$ restoring while remaining positive everywhere and exponentially small once $a_1+\gamma\gg\delta_{\mathrm{cone}}$. For wavevectors $\khat$ aligned with the eigenvectors of a frozen background $\A_0$ (including $\khat=\bm q_1$), this term shifts the transverse factor in Eq.~\eqref{eq:disp} from $\sigma+\lambda^{-1}+\kappa k^2$ to $\sigma+\lambda^{-1}+\kappa k^2-g_{\mathrm{cone}}^{[1]}(a_1,a_2)$, where the nonpositive first divided difference equals $[g_{\mathrm{cone}}(a_1)-g_{\mathrm{cone}}(a_2)]/(a_1-a_2)\leq 0$ for $a_1\neq a_2$ and $g_{\mathrm{cone}}'(a_1)\leq 0$ for $a_1=a_2$. The value $-\beta/(1-\beta)$ therefore serves as the unmodified Oldroyd--B reference threshold when interpreting finite-rate excursions. Piecewise overflow-safe evaluation of $g_{\mathrm{cone}}$ and the second-order mesoscopic source stencil are detailed in Appendix~\ref{app:intervention}.
\end{remark}

\subsection{Four-roll-mill configuration, field diagnostics, and stopping criteria}
\label{sec:stopping}

Coupled nonlinear calculations are driven in $\Omega=[0,L_b)^2$ from rest ($\bm u=\bm 0$, $\A=\bm I$) on an $N\times N$ uniform lattice ($\Delta x=L_b/N$) by the four-roll-mill body force \cite{TS2007}
\begin{equation}
\bm F_b=2s_F F_c(\sin X\cos Y,\,-\cos X\sin Y),
\qquad
(X,Y)=(k_bx,k_by),
\label{eq:four-roll-force}
\end{equation}
where $F_c=\rho\nu_s U_c k_b^2$ is the solvent-viscous Stokes force scale, $s_F$ is a dimensionless forcing multiplier, and $2s_F F_c$ is the dimensional component amplitude at position $\bm x=(x,y)$. The resulting dimensionless forcing amplitude is $F^*=2s_F F_c\ell_c/(\rho U_c^2)=2\beta s_F/\mathrm{Re}$, with $s_F=1$ at the $\beta=2/3$ baseline ($\mathrm{Re}=1$, $\mathrm{Sc}=10^5$, default $N=257$) and $s_F=6$ when matching $F^*=4/3$ at $\beta=1/9$ in Section~\ref{sec:lowbeta}. Spatial indefiniteness and local threshold exceedance are quantified across $\Omega$ by the negative-eigenvalue node fraction and domain-maximum exceedance,
\begin{equation}
\phi_-(t)=\frac{|\{\bm x\in\Omega:\lmin(\A(\bm x,t))<0\}|}{|\Omega|},
\qquad
\chi_{\max}(t)=\max_{\bm x\in\Omega}\chi(\A(\bm x,t)),
\label{eq:field-diagnostics}
\end{equation}
where $|\cdot|$ counts lattice nodes ($|\Omega|=N^2$), strict inequality $\lmin<0$ excludes zero eigenvalues, and $\chi$ is defined in Eq.~\eqref{eq:Xi}. For $\kappa>0$, the ratio $R_\Delta=\Delta x/\ell_d$ compares $\Delta x$ against the diffusion--relaxation length $\ell_d=\sqrt{\kappa\lambda}$ when interpreting the $(N,\mathrm{Sc})$ field structure in Section~\ref{sec:anatomy}. At the central extensional stagnation point (subscript $c$) of $\bm u=(u_x,u_y)$, we record
\begin{equation}
\Wi_x=\lambda\left.\frac{\partial u_x}{\partial x}\right|_c,\qquad
\Wi_{\mathrm{pub}}=\frac{\lambda}{2}
\left.\left(\frac{\partial u_x}{\partial x}
-\frac{\partial u_y}{\partial y}\right)\right|_c,
\label{eq:wi-comparison}
\end{equation}
using $\Wi_x$ in our time-series diagnostics; in the divergence-free limit, $\Wi_x$ coincides with the symmetric measure $\Wi_{\mathrm{pub}}$ used in \cite{Yu2026}.

A coupled run is stopped at a scheduled coarse diagnostic step ($\Delta t^*\approx 1.0$) whenever velocity or conformation increments become NaN or the dimensionless peak speed satisfies $\max_{\bm x}|\bm u|/U_c>50$ ($\mathrm{Ma}>0.5$), bracketing the coarse-step exit within $(t_L,t_U]$. In the eigenvalue-floor, spectral-restoring, and low-solvent series, non-intrusive diagnostics inspect every reconstructed nodal tensor at every time step without altering that coarse stopping rule, logging the exact first crossing of $\lmin<0$, $\lmin<-\beta/(1-\beta)$, and $\max_{\bm x}|\bm u|/U_c>50$. To distinguish bounded continuation up to $t^*=120$ from steady-state convergence, we evaluate the final $20\%$ of recorded finite $\Wi_x$ samples (at least $5$ samples, $n\geq 5$): a run passes tail stationarity only if its relative standard deviation is at most $5\times 10^{-3}$ and its linear drift magnitude normalized by the tail mean is at most $10^{-2}$.

%==============================================================================
\section{Verification of local dispersion predictions and smooth-flow operator consistency}
\label{sec:verify}

Before analyzing steep extensional layers in the four-roll mill, we verify the local linear predictions of Section~\ref{sec:theory} and the smooth-flow consistency of the operators in Section~\ref{sec:methods}. Specifically, we examine frozen-mode growth rates in the incompressible pseudospectral solver, transient perturbation amplification on relaxing homogeneous backgrounds in the lattice Boltzmann solver, and steady error convergence on diffusive Kolmogorov flow.

\subsection{Pseudospectral verification of threshold, short-wave plateau, and diffusive cutoff}

\begin{figure}[htbp]
\centering
\includegraphics[width=0.99\textwidth]{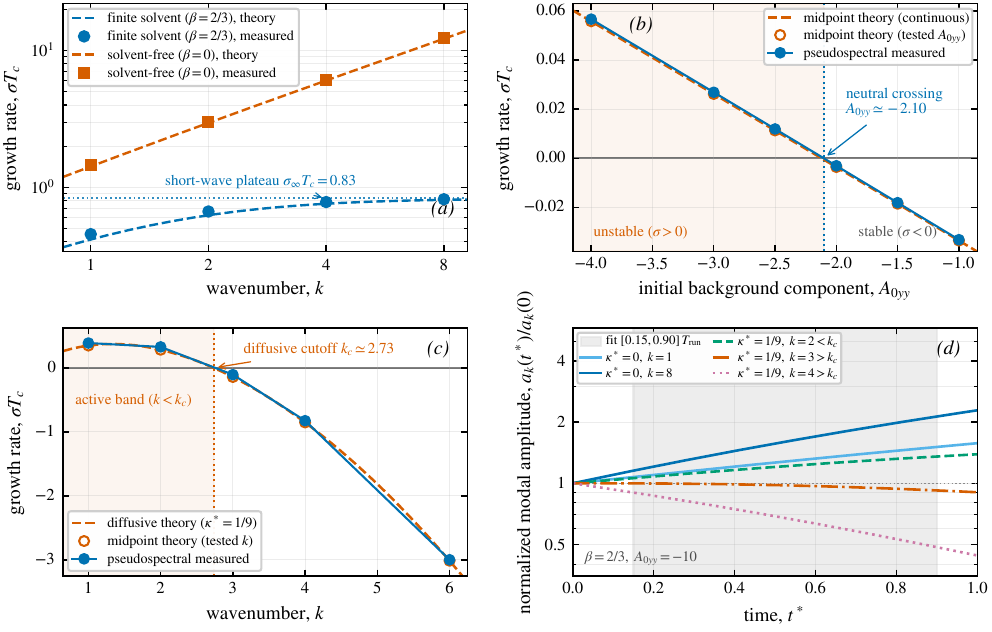}
\caption{Pseudospectral tests of the local frozen-state predictions at $\mathrm{Re}=1$ and $N=64$, with growth rates reported as $\sigma T_c$ (protocol in Table~\ref{tab:local-protocol}). Panel (b) uses $\Wi=16$ and $\bm k=(0,4)$; panels (a), (c), and (d) use $\Wi=4$ (for $\beta=\kappa=0$, $\mathrm{Sc}$ is undefined). Filled markers denote fitted numerical rates and dashed curves (with open markers in (b,c)) show analytical predictions from Eq.~\eqref{eq:disp} at the fitting-window midpoint: (a) finite-solvent plateau ($\beta=2/3$) versus linear high-$k$ growth in the solvent-free limit ($\beta=0$); (b) interpolated neutral crossing at $A_{0yy}\simeq -2.10$ versus midpoint prediction $-2.12$; (c) diffusive sign reversal between $k=2$ and $k=3$ bracketing $k_c\simeq 2.73$; and (d) normalized amplitude histories $a_k(t^*)/a_k(0)$ across the fitting window $[0.15,0.9]T_{\mathrm{run}}$ (shaded).}
\label{fig:local-validation}
\end{figure}

Incompressible pseudospectral perturbation tests reproduce the instability threshold, bounded high-wavenumber plateau, and diffusive wavenumber cutoff derived in Section~\ref{sec:theory} (Figure~\ref{fig:local-validation}). Seeding transverse eigenmodes about the homogeneous background $\bm u=\bm 0$ and $\A_0=\operatorname{diag}(5,A_{0yy})$ at $\mathrm{Re}=1$ on an $N=64$ periodic grid with wavevectors $\bm k=(0,k)$, we fit the exponential slope of the transverse velocity mode $a_k(t)=(2/N^2)|\widehat u_x(0,k,t)|$ over $[0.15,0.9]T_{\mathrm{run}}$ and compare it against Eq.~\eqref{eq:disp} evaluated at the midpoint of the relaxing background (Appendix~\ref{app:methods}, Table~\ref{tab:local-protocol}). With positive solvent viscosity ($\beta=2/3$, $\Wi=4$, $A_{0yy}=-10$), homogeneous relaxation via Eq.~\eqref{eq:relaxing-background} raises the background component to $\overline A_{yy}\simeq -8.65$ at the window midpoint, shifting the asymptotic plateau of Eq.~\eqref{eq:rate} from $1.00$ at $t=0$ to $\sigma_\infty T_c=0.831$. The measured growth rate approaches the bounded high-wavenumber plateau of Proposition~\ref{prop:threshold}, reaching $\sigma T_c=0.818$ at $k=8$ versus the midpoint-frozen prediction $0.811$ (Figure~\ref{fig:local-validation}a). In the solvent-free limit ($\beta=0$, $\kappa=0$), linear-in-$k$ Maxwell divergence reappears as predicted by Corollary~\ref{cor:ucm}, yielding $\sigma T_c=12.27$ at $k=8$ against $12.21$ predicted.

A threshold scan over $A_{0yy}$ at $\Wi=16$ and $k=4$ locates the interpolated neutral stability crossing at $A_{0yy}\simeq -2.10$, matching the initial value $-2.12$ whose window-midpoint state crosses the frozen threshold $-2$ under Eq.~\eqref{eq:relaxing-background} (Figure~\ref{fig:local-validation}b). When stress diffusion is retained ($\kappa^*=1/9$ at $\Wi=4$, $\beta=2/3$, $A_{0yy}=-10$), $\sigma T_c$ reverses sign between $k=2$ and $k=3$, bracketing the midpoint diffusive cutoff $k_c\simeq 2.73$ predicted by Eq.~\eqref{eq:cutoff} (Figure~\ref{fig:local-validation}c,d).

\subsection{Transient perturbation growth on a relaxing background}

\begin{figure}[htbp]
\centering
\includegraphics[width=0.99\textwidth]{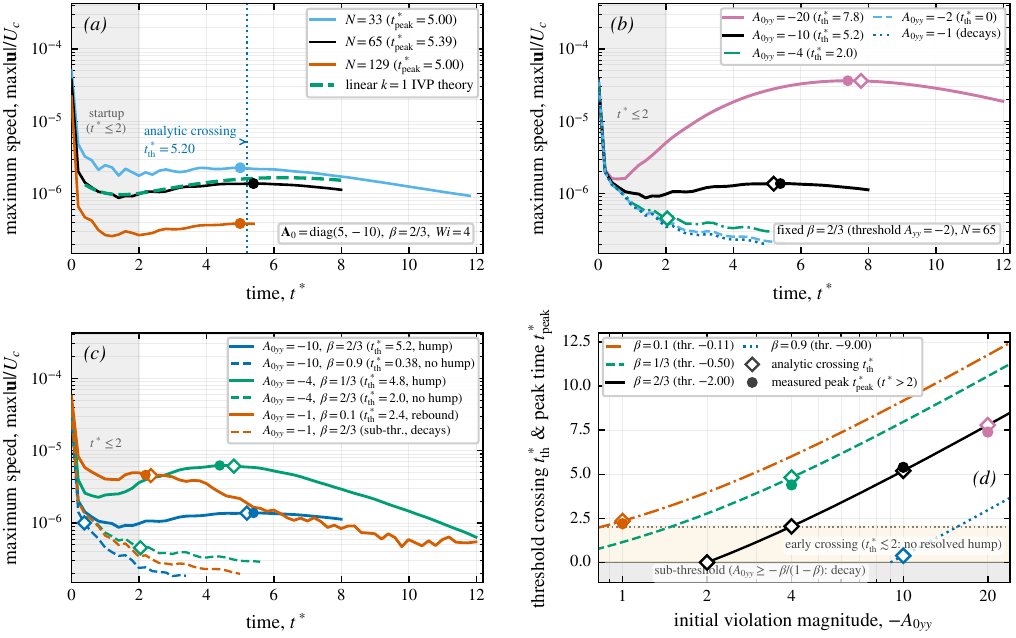}
\caption{Transient perturbation response on relaxing outside-cone backgrounds in the lattice Boltzmann solver at $\Wi=4$. Filled circles mark measured peak times $t_{\mathrm{peak}}^*$ over $t^*>2$, and open diamonds mark analytical background threshold-crossing times $t_{\mathrm{th}}^*$. Panels show: (a) grid sequence $N=33,65,129$ at $\A_0=\operatorname{diag}(5,-10)$ and $\beta=2/3$, alongside the nondiffusive non-autonomous $k=1$ linear reference initialized at $t^*=0.4$ with $(v,\zeta)=(1,0)$ and normalized at $t^*=1$ (peaking near $6.26$ convective units; LBM traces stop when the velocity steady-state tolerance triggers, e.g.\ at $t^*\simeq 5.40$ on $N=129$); (b) initial eigenvalue scan at $\beta=2/3$ ($N=65$, threshold $-2$); (c) paired solvent-ratio histories at fixed $\A_0$ ($N=65$), contrasting lower-$\beta$ transient humps (solid) with higher-$\beta$ decay (dashed); and (d) analytical $t_{\mathrm{th}}^*(-A_{0yy};\beta)$ compared against measured $t_{\mathrm{peak}}^*$ (for $\beta=0.1$ and $A_{0yy}=-1$, the $t^*>2$ marker records the window-left maximum at $t^*\simeq 2.20$ following the rebound near $t^*\simeq 1.8$).}
\label{fig:growth}
\end{figure}

In the lattice Boltzmann solver at $\Wi=4$, seeding $\A_0=\operatorname{diag}(5,A_{0yy})$ with broadband velocity noise of nominal amplitude $10^{-6}$ tracks $a_{\mathrm{LBM}}(t)=\max_{\Omega}|\bm u|/U_c$ as the homogeneous background relaxes via Eq.~\eqref{eq:relaxing-background}. At $\beta=2/3$ (threshold $-2$), every state with $A_{0yy}<0$ satisfies $\det\A_0<0$, yet subcritical and threshold states $A_{0yy}=-1$ and $-2$ decay across the sampled window without a growth hump, whereas deeply supercritical states $A_{0yy}=-10$ and $-20$ develop pronounced transient humps before decaying (Figure~\ref{fig:growth}b). The weakly supercritical case $A_{0yy}=-4$, whose supercritical interval ends near $t_{\mathrm{th}}^*\simeq 2.04$, shows no resolved post-startup hump.

Along Eq.~\eqref{eq:relaxing-background}, an initially supercritical component $A_{0yy}<-1/r$ crosses back through the frozen-state threshold $-1/r$ at the analytical background threshold-crossing time
\[
t_{\mathrm{th}}=\lambda\ln\!\left[\frac{r(1-A_{0yy})}{r+1}\right],
\qquad
r=\frac{\mu_p}{\mu_s}=\frac{1-\beta}{\beta}.
\]
For $t_{\mathrm{th}}^*=t_{\mathrm{th}}/T_c$ at $\beta=2/3$ ($r=1/2$) and $\Wi=4$, $A_{0yy}=-10$ gives $t_{\mathrm{th}}^*=5.2$, compared with measured perturbation peak times $t_{\mathrm{peak}}^*\in[5.0,5.4]$ over $t^*>2$ across $N=33,65,129$; $A_{0yy}=-20$ gives $t_{\mathrm{th}}^*=7.8$ versus the measured peak at $t^*\simeq 7.4$ (Figure~\ref{fig:growth}a,d). A shift in $\beta$ at fixed $\A_0$ alters both $-1/r$ and $t_{\mathrm{th}}^*$: $\operatorname{diag}(5,-1)$ decays at $\beta=2/3$ but grows transiently at $\beta=0.1$, whereas $\operatorname{diag}(5,-10)$ exhibits a clear hump at $\beta=2/3$ but none at $\beta=0.9$, where $-1/r=-9$ leaves only a brief supercritical interval (Figure~\ref{fig:growth}c). At $\operatorname{diag}(5,-10)$, the three grids $N=33,65,129$ yield closely matching curve shapes and peak times over $t^*>2$ under grid-dependent deterministic linear-congruential velocity perturbations (Figure~\ref{fig:growth}a, which also plots the non-autonomous $k=1$ linear reference peaking near $6.26$). Because these broadband perturbations are not mode-matched across grids, this three-grid agreement serves as a qualitative check on peak-time consistency, whereas exact-solution convergence orders are measured on steady Kolmogorov flow in Section~\ref{sec:kolmogorov}.

\subsection{Smooth-flow convergence and operator consistency on Kolmogorov flow}
\label{sec:kolmogorov}

Coupled accuracy and smooth-flow operator consistency are verified on the steady Kolmogorov flow driven by $\bm F_b=F_k(\sin(k_fy),0)$ with $k_f=k_b$ and $F_k=(\mu_s+\mu_p)k_f^2 U_c$ at $\beta=2/3$, $\mathrm{Re}=1$, $\Wi=8$, and $\mathrm{Sc}=10^5$ on $N=33,65,129$ \cite{BB2010}. With stress diffusion retained, the exact steady velocity profile is $u_\kappa=U_\kappa\sin(k_fy)$ with $U_\kappa=F_k/[k_f^2(\mu_s+\mu_p/(1+\delta))]$ and $\delta=\kappa\lambda k_f^2=5.33\times 10^{-5}$, corresponding to a relative amplitude shift $U_\kappa/U_c-1=1.78\times 10^{-5}$ (Appendix~\ref{app:methods}). Under acoustic refinement ($\Delta t\propto\Delta x$ at fixed $\mathrm{Ma}=10^{-2}$), relative discrete $L^2$ streamwise-velocity errors at $t^*\simeq 60$ against the $t=\infty$ steady profile $u_\kappa$ under SM-1 are $3.03\times 10^{-3}$, $7.82\times 10^{-4}$, and $2.00\times 10^{-4}$ across $N=33,65,129$ (orders $2.00$ and $1.99$). Measured against the exact time-dependent harmonic solution from rest at the snapshot time $t^*\simeq 60$ ($7.5\lambda$, where $A_{xx}$ still carries a $1.18\times 10^{-3}$ physical relaxation tail; Appendix~\ref{app:methods}), the coupled relative $L^2$ errors across $N=33,65,129$ are $3.03\times 10^{-3}$, $7.79\times 10^{-4}$, and $1.98\times 10^{-4}$ in $u_x$, $3.02\times 10^{-3}$, $7.79\times 10^{-4}$, and $1.98\times 10^{-4}$ in $A_{xy}$, and $6.02\times 10^{-3}$, $1.55\times 10^{-3}$, and $3.94\times 10^{-4}$ in $A_{xx}$, yielding pairwise convergence orders of $2.00$ in all three fields ($A_{yy}=1$ identically). Because this smooth solution remains uniformly inside $\mathcal K^{\circ}$ ($\min_{t^*,\bm x}\lmin(\A)\simeq 0.172$--$0.173$ over $t^*\in[0,60]$, settling to $0.501$ at $t^*\simeq 60$), the shifted floor $\Pi_{-\gamma+\varepsilon}$ ($c_\gamma=0.5$) is never triggered ($0$ projections), leaving all velocity and conformation fields bit-identical across all $17$ double-precision digits, while $\bm S_{\mathrm{cone}}(\A)$ ($c_\gamma=0.5$, $r_{\mathrm{cone}}=0.01$) alters the coupled relative $L^2$ errors by at most $2.7\times 10^{-12}$ in $u_x$ and $7.3\times 10^{-12}$ across $\A$, leaving all convergence orders unchanged and confirming that both operators remain non-intrusive in resolved in-cone flow.

%==============================================================================
\section{Stress--momentum coupling, indefinite field structure, and finite-horizon stabilization in the four-roll mill}
\label{sec:causal}

In discrete four-roll-mill computations, steep extensional layers drive $\A$ outside $\mathcal K$, testing how discrete stress--momentum coupling and threshold-anchored spectral control govern finite-horizon behavior. Sections~\ref{sec:cure} and~\ref{sec:clipping} contrast the five momentum-coupling schemes and step-resolved intervention histories at moderate solvent viscosity ($\beta=2/3$, $\Wi=50$). Section~\ref{sec:anatomy} resolves the spatial structure and two-dimensional determinant balance of persistent outside-cone layers under unmodified SM-1 at $\Wi=20$, and Sections~\ref{sec:lowbeta} and~\ref{sec:shifted-cone} isolate parameter scaling and finite-horizon stabilization at low solvent viscosity ($\beta=1/9$, $\Wi=50$).

\subsection{Momentum-coupling survival versus nodal constraint enforcement at moderate solvent viscosity}

\subsubsection{Five momentum-coupling schemes and feedback-gain sensitivity}
\label{sec:cure}

At $\Wi=50$, $\beta=2/3$ (frozen-state reference threshold $-2$), $N=257$, $\mathrm{Sc}=10^5$, and $\mathrm{Re}=1$, the five momentum-coupling schemes of Section~\ref{sec:coupling} divide sharply in finite-horizon survival (Table~\ref{tab:matrix}). Unmodified FD-TC terminates in $(46.0,47.0]$, whereas FD-Guo and all three local stress-moment schemes (SM-1, SM-Guo, and SM-TC) reach $t^*=120$ despite strongly indefinite conformation fields with $\min_{\bm x}\lmin\simeq -147.9$ to $-150.5$ and $\phi_-\simeq 2.44\%$--$2.57\%$. Threshold-anchored spectral control reinforces this separation: SM-1 reaches $t^*=120$ under both the shifted floor $\Pi_{-1+\varepsilon}$ ($c_\gamma=0.5$) and the smooth restoring source $\bm S_{\mathrm{cone}}(\A)$ ($c_\gamma=0.5$, $r_{\mathrm{cone}}=0.01$), whereas FD-TC terminates near $t^*\simeq 61$ under strict-SPD projection, shifted-cone projection, and smooth spectral restoring alike. All six runs reaching $t^*=120$ in Table~\ref{tab:matrix} remain unsteady and fail the tail-stationarity criterion of Section~\ref{sec:stopping}.

\begin{table}[htbp]
\centering\linespread{0.90}\selectfont\scriptsize
\setlength{\tabcolsep}{3pt}
\renewcommand{\arraystretch}{1.02}
\setlength{\abovecaptionskip}{4pt}
\begin{tabular}{@{}>{\raggedright\arraybackslash}p{\dimexpr0.19\textwidth-2\tabcolsep\relax}>{\raggedright\arraybackslash}p{\dimexpr0.14\textwidth-2\tabcolsep\relax}>{\raggedright\arraybackslash}p{\dimexpr0.27\textwidth-2\tabcolsep\relax}>{\raggedright\arraybackslash}p{\dimexpr0.18\textwidth-2\tabcolsep\relax}>{\raggedright\arraybackslash}p{\dimexpr0.13\textwidth-2\tabcolsep\relax}>{\raggedright\arraybackslash}p{\dimexpr0.09\textwidth-2\tabcolsep\relax}@{}}
\toprule
Scheme / intervention & Polymer feedback & Background force / modification & Outcome ($t^*\leq 120$)
& Final $\min_{\bm x}\lmin$ & Final $\phi_-$\\
\midrule
\multicolumn{6}{@{}l}{\emph{Five momentum-coupling schemes (unmodified conformation state, $g=1.00$)}}\\
FD-TC  & Force divergence & Time-consistent & Terminated in $(46.0,47.0]$ & -- & --\\
FD-Guo & Force divergence & Second-order Guo & Reaches $t^*=120$ & $-147.9$ & $2.44\%$\\
SM-1   & Stress moment    & First-order forcing & Reaches $t^*=120$ & $-150.5$ & $2.57\%$\\
SM-Guo & Stress moment    & Second-order Guo & Reaches $t^*=120$ & $-150.5$ & $2.57\%$\\
SM-TC  & Stress moment    & Time-consistent & Reaches $t^*=120$ & $-150.5$ & $2.57\%$\\
\midrule
\multicolumn{6}{@{}l}{\emph{FD-TC and paired SM-1 state/source interventions ($g=1.00$)}}\\
FD-TC + conformation filter & Force divergence & Order-$8$ filter on $\A$ ($\sigma_f=0.2$) & Terminated in $(20.0,21.0]$ & -- & --\\
FD-TC + one-sided floor & Force divergence & Target $a_1\geq 10^{-8}$ only (non-SPD from $t^*=5.65$) & Terminated in $(61.0,62.0]$ & -- & --\\
FD-TC + two-eigenvalue floor & Force divergence & Target $a_1,a_2\geq 10^{-8}$ (SPD to $t^*=61.2406$) & Terminated in $(61.0,62.0]$ & -- & --\\
FD-TC + shifted floor & Force divergence & Target $a_1,a_2\geq -1+10^{-8}$ ($c_\gamma=0.5$) & Terminated in $(60.0,61.0]$ & -- & --\\
FD-TC + restoring source & Force divergence & Smooth $\bm S_{\mathrm{cone}}$ ($c_\gamma=0.5$, $r_{\mathrm{cone}}=0.01$) & Terminated in $(60.0,61.0]$ & -- & --\\
SM-1 + shifted floor & Stress moment & Target $a_1,a_2\geq -1+10^{-8}$ ($c_\gamma=0.5$) & Reaches $t^*=120$ & $-1.00$ & $2.41\%$\\
SM-1 + restoring source & Stress moment & Smooth $\bm S_{\mathrm{cone}}$ ($c_\gamma=0.5$, $r_{\mathrm{cone}}=0.01$) & Reaches $t^*=120$ & $-1.35$ & $2.24\%$\\
\midrule
\multicolumn{6}{@{}l}{\emph{FD-TC polymer-force gain scan ($g\bm F_p$, unmodified conformation state)}}\\
FD-TC ($g=0.50$) & Scaled force div.\ ($0.50\bm F_p$) & Time-consistent & Terminated in $(43.0,44.0]$ & -- & --\\
FD-TC ($g=0.70$) & Scaled force div.\ ($0.70\bm F_p$) & Time-consistent & Terminated in $(44.0,45.0]$ & -- & --\\
FD-TC ($g=0.85$) & Scaled force div.\ ($0.85\bm F_p$) & Time-consistent & Terminated in $(45.0,46.0]$ & -- & --\\
FD-TC ($g=1.00$, baseline) & Scaled force div.\ ($1.00\bm F_p$) & Time-consistent & Terminated in $(46.0,47.0]$ & -- & --\\
FD-TC ($g=1.15$) & Scaled force div.\ ($1.15\bm F_p$) & Time-consistent & Terminated in $(47.0,48.0]$ & -- & --\\
\bottomrule
\end{tabular}
\caption{Momentum-coupling comparison, paired FD-TC and SM-1 spectral interventions, and FD-TC polymer-force gain scan at nominal $\Wi=50$ ($257^2$, $\mathrm{Sc}=10^5$, $\beta=2/3$, $\mathrm{Re}=1$; formulations in Section~\ref{sec:methods} and Appendix~\ref{app:intervention}, step-resolved FD-TC event sequences in Table~\ref{tab:clip-timeline}). All six runs reaching $t^*=120$ fail the tail-stationarity test of Section~\ref{sec:stopping}. Dashes indicate that terminal conformation metrics were either unsaved after coarse-step exit or nonfinite, and do not imply positive definiteness.}
\label{tab:matrix}
\end{table}

Within the stress-moment family, SM-1 \cite{Yu2026}, SM-Guo, and SM-TC share the local second-moment source $T_i$ of Eq.~\eqref{eq:Ti} and differ solely in the discretization of $\bm F_b$: all three reach $t^*=120$ with final $\Wi_x$ spanning only $1.50\times 10^{-5}$ and virtually identical outside-cone metrics. Within the force-divergence family, omitting $R_i(\bm F_b+\bm F_p)$ and applying the half-force velocity shift in FD-Guo avoids the early exit of FD-TC, while switching from FD-TC to SM-TC simultaneously replaces centered-difference $\nabla\!\cdot\!\bm\sigma_p$ with $T_i$ and removes $\bm F_p$ from $R_i$ in Eq.~\eqref{eq:forcing-kernels}. Across the polymer-force gain scan $g\bm F_p$ with $g\in\{0.50,0.70,0.85,1.00,1.15\}$ in FD-TC, the upper endpoint of the coarse exit bracket moves monotonically from $t^*=44.0$ at $g=0.50$ to $48.0$ at $g=1.15$, spanning only $4T_c$. Varying polymer-force amplitude within FD-TC therefore shifts exit timing slightly without reproducing SM-TC continuation.

\subsubsection{Failure mechanism under force-divergence coupling}
\label{sec:clipping}

Every-step diagnostics clarify why nodal constraint enforcement fails to prevent termination under FD-TC at $\Wi=50$ and $\beta=2/3$ (Table~\ref{tab:clip-timeline}). Under the strict-SPD two-eigenvalue floor ($c_\gamma=0$, $\ell_{\mathrm{floor}}=10^{-8}$), the second eigenvalue first requires projection at $t^*=5.65$. Every reconstructed nodal tensor supplying $\bm\sigma_p$ then remains finite and strictly SPD through the first crossing of the velocity stopping threshold $U_{\max}=\max_{\bm x}|\bm u|/U_c>50$ at $t^*=61.2406$ ($U_{\max}=50.8$, $\min\lmin^{\mathrm{post}}=9.54\times 10^{-9}>0$ over $[0,61.2406]$; Appendix~\ref{app:revision-checks}). Over $t^*=47.0\to 61.0$, scheduled $\Delta t^*\approx 1$ samples remain well within the low-Mach regime ($U_{\max}\leq 0.79$, $\mathrm{Ma}<0.008$), yet positive polymer stretch and shear stress already escalate inside $\mathcal K^{\circ}$: $\max_{\bm x}\operatorname{tr}\A$ rises from $4.52\times 10^3$ to $8.13\times 10^4$ and $\max_{\bm x}A_{xy}$ from $3.03\times 10^2$ to $1.92\times 10^4$ (whereas both stay near $6.0\times 10^3$ and $1.3\times 10^2$ in FD-Guo and SM-TC). This positive-cone runaway drives peak speed past $U_{\max}=50.8$ ($\mathrm{Ma}\simeq 0.508$, violating low-Mach lattice Boltzmann scaling), whereas non-SPD reconstructed tensors and nonfinite conformation moments appear only $33$ and $124$ steps later ($t^*=61.2453$ and $61.2581$) before coarse exit at $t^*=62.0$.

\begin{table}[htbp]
\centering\linespread{0.82}\selectfont
\footnotesize
\setlength{\tabcolsep}{3pt}
\renewcommand{\arraystretch}{0.90}
\setlength{\abovecaptionskip}{4pt}
\begin{tabular}{@{}>{\raggedright\arraybackslash}p{\dimexpr0.19\textwidth-2\tabcolsep\relax}>{\raggedright\arraybackslash}p{\dimexpr0.25\textwidth-2\tabcolsep\relax}>{\raggedleft\arraybackslash}p{\dimexpr0.10\textwidth-2\tabcolsep\relax}>{\raggedleft\arraybackslash}p{\dimexpr0.14\textwidth-2\tabcolsep\relax}>{\raggedright\arraybackslash}p{\dimexpr0.32\textwidth-2\tabcolsep\relax}@{}}
\toprule
Eigenvalue-floor operator & Recorded event in FD-TC & Step & $t^*$ & Reconstructed nodal diagnostic \\
\midrule
\parbox[t][0pt][t]{\linewidth}{\raggedright One-sided floor (target $a_1\geq\varepsilon$ only)}
  & First non-SPD reconstructed nodal output & $40043$ & $5.65$ & $\lmin^{\mathrm{post}}=-3.92\times10^{-5}$ (both input eigenvalues $<0$) \\
  & Coarse diagnostic termination bracket & -- & $(61.0,62.0]$ & Stopped by coarse diagnostic rule \\
\midrule
\parbox[t][0pt][t]{\linewidth}{\raggedright Two-eigenvalue floor (target $a_1,a_2\geq\varepsilon$)}
  & First active second-eigenvalue floor & $40043$ & $5.65$ & $\lmin^{\mathrm{post}}=1.00\times10^{-8}$ (strictly SPD) \\
  & First velocity-threshold crossing ($U_{\max}>50$) & $433864$ & $61.2406$ & $U_{\max}=50.8$ ($\mathrm{Ma}\simeq0.508$); $\min\lmin^{\mathrm{post}}=9.54\times10^{-9}$ on $[0,61.2406]$ \\
  & First non-SPD reconstructed nodal tensor & $433897$ & $61.2453$ & $+33$ steps after $U_{\max}>50$ crossing \\
  & First nonfinite conformation moment & $433988$ & $61.2581$ & $+124$ steps after $U_{\max}>50$ crossing \\
  & Coarse diagnostic rule exit & $439270$ & $62.0$ & Coarse termination bracket $(61.0,62.0]$ \\
\midrule
\parbox[t][0pt][t]{\linewidth}{\raggedright Shifted two-eigenvalue floor ($c_\gamma=0.5$, target $a_1,a_2\geq -1+\varepsilon$)}
  & First negative reconstructed nodal eigenvalue & $32335$ & $4.56$ & $\lmin^{\mathrm{post}}=-1.98\times10^{-6}$ ($\phi_-=0.003\%$) \\
  & First velocity-threshold crossing ($U_{\max}>50$) & $428930$ & $60.5442$ & $U_{\max}=51.8$ ($\mathrm{Ma}\simeq0.518$); $\min\lmin^{\mathrm{post}}=-1+9.42\times10^{-9}$ on $[0,60.5442]$ \\
  & First reconstructed tensor below $-2$ & $428976$ & $60.5507$ & $+46$ steps after $U_{\max}>50$ crossing \\
  & First nonfinite conformation moment & $429059$ & $60.5624$ & $+129$ steps after $U_{\max}>50$ crossing \\
  & Coarse diagnostic rule exit & $432185$ & $61.0$ & Coarse termination bracket $(60.0,61.0]$ \\
\bottomrule
\end{tabular}
\caption{Step-resolved event sequence under the one-sided, strict-SPD ($c_\gamma=0$), and shifted ($c_\gamma=0.5$) eigenvalue floors ($\varepsilon=10^{-8}$) in FD-TC at $257^2$, nominal $\Wi=50$, $\mathrm{Sc}=10^5$, $\beta=2/3$, and $\mathrm{Re}=1$ ($U_{\max}=\max_{\bm x}|\bm u|/U_c$). Through the first crossing of $U_{\max}>50$, every reconstructed nodal tensor supplying polymer-stress feedback stays finite and strictly SPD under $c_\gamma=0$ ($\min\lmin^{\mathrm{post}}=9.54\times 10^{-9}>0$, $4.6\times 10^{-10}$ below $\varepsilon$) and keeps $\lmin^{\mathrm{post}}>-1$ (strictly above the reference threshold $-2$) under $c_\gamma=0.5$ ($\min\lmin^{\mathrm{post}}=-1+9.42\times 10^{-9}$, $5.8\times 10^{-10}$ below $\ell_{\mathrm{floor}}=-1+\varepsilon$ from finite-precision $2\times 2$ spectral assembly conditioning; Appendix~\ref{app:revision-checks}).}
\label{tab:clip-timeline}
\end{table}

Four auxiliary conformation interventions confirm that one-sided or sub-threshold floors, kink-free spectral restoring, and linear high-wavenumber filtering likewise fail to prevent FD-TC breakdown (Tables~\ref{tab:matrix} and~\ref{tab:clip-timeline}). First, a one-sided floor bounding only $a_1$ leaves the reconstructed tensor indefinite as early as $t^*=5.65$ once both pre-projection eigenvalues turn negative, yet exits in the same coarse interval $(61.0,62.0]$ as the strict-SPD floor. Second, under the shifted floor $\Pi_{-1+\varepsilon}$ ($c_\gamma=0.5$), negative eigenvalues enter at $t^*=4.56$, while every conformation eigenvalue supplying polymer stress stays strictly above $-1$ (well above the reference threshold $-2$) up to the velocity-threshold crossing at $t^*=60.5442$ ($U_{\max}=51.8$, coarse exit $(60.0,61.0]$). Third, replacing hard projection with the clip-free smooth source $\bm S_{\mathrm{cone}}(\A)$ ($c_\gamma=0.5$, $r_{\mathrm{cone}}=0.01$, $0$ clips) removes projection kinks altogether, yet FD-TC still crosses $U_{\max}>50$ at $t^*=60.6297$. Fourth, applying the order-$8$ spatial conformation filter at every step triggers exit even earlier in $(20.0,21.0]$. FD-TC breakdown therefore requires neither projection kinks nor pre-event excursions below $-2$. By elimination, it originates in discrete stress--momentum coupling across steep one- to two-cell stress layers, where the tested schemes differ in forcing structure: FD-TC combines unshifted centered-difference polymer forcing $\bm F_p=\nabla\!\cdot\!\bm\sigma_p$ with the quadratic population correction $R_i(\bm F_b+\bm F_p)$ of Eq.~\eqref{eq:forcing-kernels} (whose second velocity moment equals $\sum_i\bm c_i\bm c_i R_i(\bm F)=\tfrac{1}{4}(\Delta t^2/\rho)\bm F\bm F^{\!\top}$), whereas replacing this stencil with half-force shifting, prefactor $1-\Omega_s/2$, and no $R_i$ term (FD-Guo) or injecting $\bm\sigma_p$ locally without spatial differentiation via $T_i$ (SM-TC) avoids positive-cone runaway.

\subsection{Spatial structure and determinant balance of persistent outside-cone layers}
\label{sec:anatomy}

\subsubsection{Sub-step localization of first cone exit and sensitivity to grid and stress diffusion}
\label{sec:first-cross}

Whereas $\Wi=50$ generates strongly unsteady strand dynamics, lowering the Weissenberg number to $\Wi=20$ at $\beta=2/3$ and $\mathrm{Re}=1$ yields coherent birefringent strands whose outside-cone structure can be tracked clearly across five $(N,\mathrm{Sc})$ combinations under unmodified SM-1. On the $N=257$, $\mathrm{Sc}=10^5$ baseline initialized at $\A=\bm I$, five-stage sub-step diagnostics record the first sign change of both $\min_{\bm x}\det\A$ and $\min_{\bm x}\lmin(\A)$ at step $35328$ ($t^*=4.99$): moments remain strictly inside $\mathcal K^{\circ}$ after collision ($\min_{\bm x}\lmin=2.87\times 10^{-5}$) but turn negative immediately after lattice streaming ($\min_{\bm x}\lmin=-3.71\times 10^{-6}$; Appendix~\ref{app:first-cross}, Table~\ref{tab:first-cross}), directly realizing the constraint-relaxation criterion of Eq.~\eqref{eq:discrete-relaxation-test}.

\begin{figure}[htbp]
\centering
\includegraphics[width=0.99\textwidth]{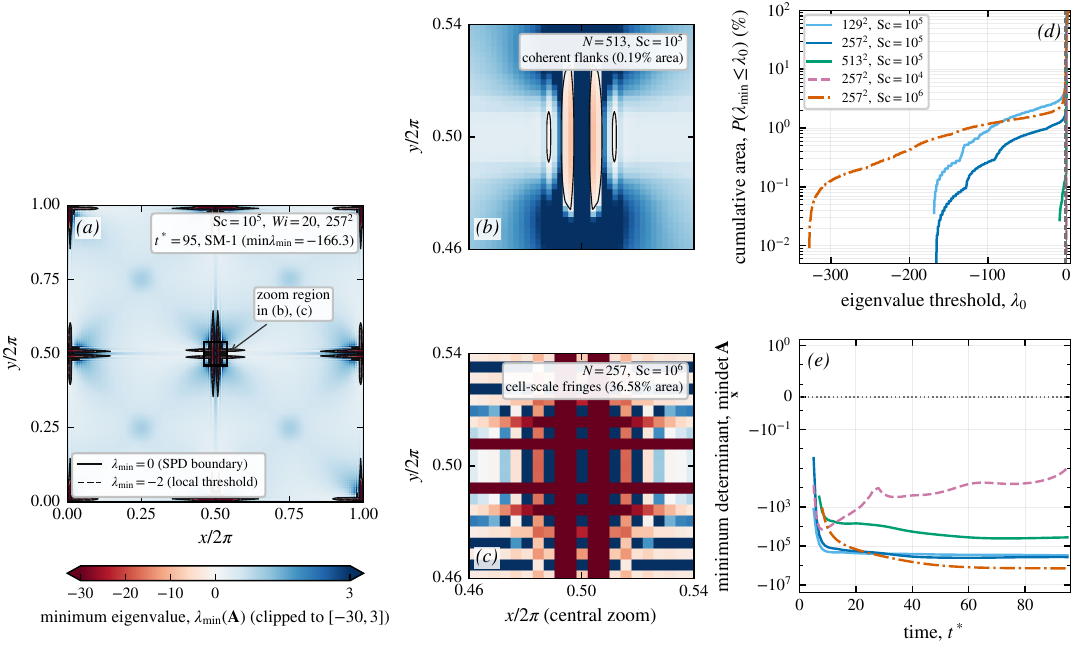}
\caption{Outside-cone conformation structure under unmodified SM-1 at $\Wi=20$, $\beta=2/3$, and $\mathrm{Re}=1$ ($t^*=95$, summary metrics in Table~\ref{tab:field-sensitivity}). Panels show: (a) $\lmin(\A)$ field on the $N=257$, $\mathrm{Sc}=10^5$ baseline (color scale clipped to $[-30,3]$, marking $\lmin=0$ solid and $\lmin=-2$ dashed); (b,c) central zoom ($x/2\pi,y/2\pi\in[0.46,0.54]$) contrasting coherent strand flanks at $N=513$, $\mathrm{Sc}=10^5$ (b) against cell-scale fringes along the connected central cross at $N=257$, $\mathrm{Sc}=10^6$ (c); (d) empirical cumulative area distribution $P(\lmin\leq\lambda_0)$ across all five $(N,\mathrm{Sc})$ runs at $t^*=95$; and (e) post-exit $\min_{\bm x}\det\A(t^*)$ histories (symmetric logarithmic scale; same legend as (d)).}
\label{fig:map}
\end{figure}

After the discrete state first leaves $\mathcal K_h$ at $t^*\approx 5$--$8$, all five $(N,\mathrm{Sc})$ runs sustain finite negative determinant minima across every subsequent sample through $t^*=95$ (Figure~\ref{fig:map}e). At $t^*=95$ on the $N=257$, $\mathrm{Sc}=10^5$ baseline, $\lmin<0$ nodes occupy $2.23\%$ of $\Omega$ ($\min_{\bm x}\lmin=-166.3$) along narrow layers only a few cells wide bordering the extensional strands (Figure~\ref{fig:map}a and Table~\ref{tab:field-sensitivity}). Grid refinement across $N=129,257,513$ at fixed $\mathrm{Sc}=10^5$ lowers the grid-to-diffusion ratio $R_\Delta=\Delta x/\ell_d$ from $4.22$ to $2.12$ and $1.06$, reducing $\phi_-$ from $6.63\%$ to $2.23\%$ and $0.19\%$; meanwhile, $\min_{\bm x}\lmin$ stays near $-168.8$ and $-166.3$ before rising sharply to $-8.98$ once $R_\Delta\approx 1$ on $N=513$ (Figure~\ref{fig:map}b,d). These unresolved layers thus remain grid-sensitive without reaching asymptotic spatial convergence. Changing $\mathrm{Sc}$ at fixed $N=257$ modifies the diffusive PDE itself: stronger diffusion at $\mathrm{Sc}=10^4$ ($R_\Delta\simeq 0.67$) confines $\lmin<0$ to $0.0076\%$ ($5$ nodes, $\min_{\bm x}\lmin=-0.0066$), whereas weaker diffusion at $\mathrm{Sc}=10^6$ ($R_\Delta\simeq 6.70$) expands $\phi_-$ to $36.58\%$ ($\min_{\bm x}\lmin=-327.8$). At $\mathrm{Sc}=10^6$, cell-scale fringes develop along the connected central cross (Figure~\ref{fig:map}c), while across the full domain $89.7\%$ of periodic four-neighbor negative-eigenvalue components ($9.4\%$ of all $\lmin<0$ nodes) fragment into isolated single cells outside the central stagnation zone.

\begin{table}[htbp]
\centering
\small
\setlength{\tabcolsep}{3pt}
\begin{tabular}{@{}rrrrrr@{}}
\toprule
$N$ & $\mathrm{Sc}$ & $\min_{\bm x}\lmin$ & $\lmin<0$ area (\%) & $\det\A<0$ area (\%) & $\lmin<-2$ area (\%) \\
\midrule
129 & $10^5$ & $-168.8$ & 6.63 & 6.25 & 4.16 \\
257 & $10^5$ & $-166.3$ & 2.23 & 2.15 & 1.56 \\
513 & $10^5$ & $-8.98$  & 0.19 & 0.19 & 0.12 \\
\midrule
257 & $10^4$ & $-0.0066$ & $<0.01$ & $<0.01$ & 0.00 \\
257 & $10^6$ & $-327.8$ & 36.58 & 35.89 & 4.25 \\
\bottomrule
\end{tabular}
\caption{Outside-cone diagnostics for unmodified SM-1 fields at $\Wi=20$, $\beta=2/3$, and $\mathrm{Re}=1$ ($t^*=95$). The top three rows vary $N$ at $\mathrm{Sc}=10^5$; the bottom two rows together with the $N=257$, $\mathrm{Sc}=10^5$ baseline vary $\mathrm{Sc}$ at $N=257$ (at $\mathrm{Sc}=10^4$, $\lmin<0$ and $\det\A<0$ each occupy $0.0076\%$, or $5$ nodes).}
\label{tab:field-sensitivity}
\end{table}

\subsubsection{Constitutive recovery and closure residual across negative-determinant cores}

\begin{figure}[htbp]
\centering
\includegraphics[width=0.98\textwidth]{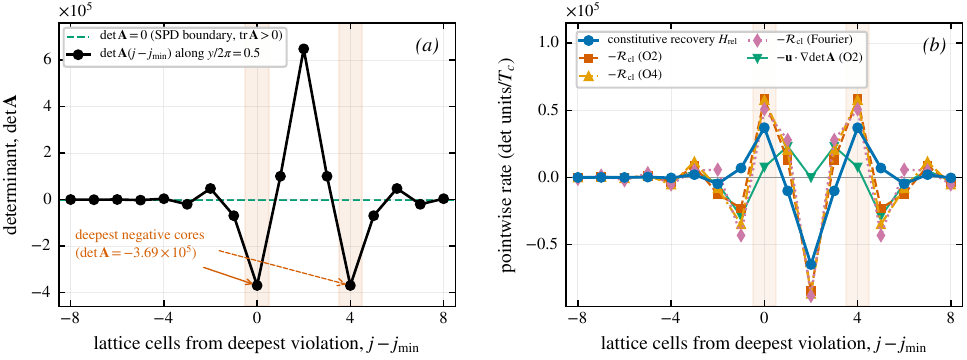}
\caption{Cross-strand profiles of the reconstructed determinant balance for the baseline SM-1 field ($\Wi=20$, $N=257$, $\mathrm{Sc}=10^5$, $t^*=95$) along $y/2\pi=0.5$ through the deepest violation ($j=j_{\min}$) using a centered three-snapshot time derivative. Panel (a) shows that $\det\A$ changes sign within one to two grid intervals across the positive strand center ($j-j_{\min}=2$) and symmetric negative cores ($j-j_{\min}=0$ and $4$, $\min\det\A=-3.69\times 10^5$). Panel (b) compares the restorative relaxation term $H_{\mathrm{rel}}$, the negative closure residual $-\mathcal R_{\mathrm{cl}}$ (plotted with a minus sign for direct magnitude comparison against $H_{\mathrm{rel}}$) across second-order (O2), fourth-order (O4), and Fourier spatial derivatives, and resolved advection $-\bm u\cdot\nabla\det\A$ (O2); subset-averaged values over $\det\A<0$ are listed in Table~\ref{tab:budget-mask}.}
\label{fig:budget}
\end{figure}

We evaluate Eq.~\eqref{eq:det2d} on the saved $t^*=95$ snapshots to test whether Oldroyd--B constitutive relaxation remains restorative inside these persistent indefinite layers. Across the $\det\A<0$ subsets of all five $(N,\mathrm{Sc})$ fields in Table~\ref{tab:field-sensitivity}, the dimensionless relaxation rate $T_c H_{\mathrm{rel}}=(\operatorname{tr}\A-2\det\A)/\Wi$ is strictly positive at $98.9\%$--$100\%$ of nodes, showing that constitutive relaxation raises the determinant across nearly the entire negative-determinant region (Section~\ref{sec:selfheal}). Along the symmetry line $y/2\pi=0.5$ on the $N=257$, $\mathrm{Sc}=10^5$ baseline (Figure~\ref{fig:budget}), $\det\A$ drops to $-3.69\times 10^5$ across one to two grid cells, where positive $H_{\mathrm{rel}}$ is offset by a negative reconstructed closure residual $\mathcal R_{\mathrm{cl}}$. Because the subset mean of $\mathcal R_{\mathrm{cl}}$ shifts substantially across second-order, fourth-order, and spectral Fourier spatial derivatives (Appendix~\ref{app:det}, Table~\ref{tab:budget-mask}), the quasi-steady core balance cannot be attributed to any single resolved term.

\FloatBarrier

\subsection{Parameter scaling and finite-horizon stabilization at low solvent viscosity}

\subsubsection{Weak-forcing survival versus baseline-forcing termination}
\label{sec:lowbeta}

\begin{figure}[htbp]
\centering
\includegraphics[width=0.98\textwidth]{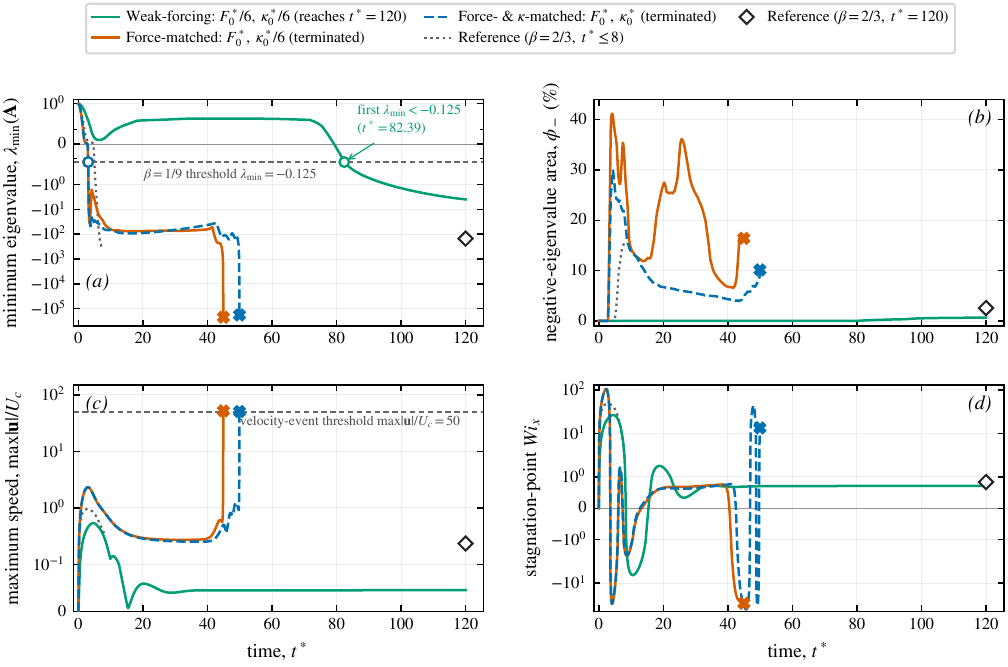}
\caption{Unmodified SM-1 histories across the low-solvent control cases of Table~\ref{tab:lowbeta-parameters} ($\beta=1/9$, $\Wi=50$, $N=257$, $\mathrm{Re}=1$): (a) minimum eigenvalue $\lmin(\A)$ (symmetric logarithmic scale, linear within $|\lmin|\leq 0.125$), (b) negative-eigenvalue node percentage $100\phi_-$, (c) dimensionless peak speed $\max_{\bm x}|\bm u|/U_c$ (symmetric logarithmic scale), and (d) stagnation-point $\Wi_x$ (symmetric logarithmic scale, linear within $|\Wi_x|\leq 1$). Curves combine $0.1T_c$ sampling with exact step-resolved first-event markers (open circles on $\lmin=-0.125$) and end at the first crossing of $\max_{\bm x}|\bm u|/U_c=50$ for the two terminating runs; $\beta=2/3$ reference markers indicate the step-monitored prefix and saved $t^*=120$ endpoint.}
\label{fig:lowbeta}
\end{figure}

At $\beta=2/3$, the sub-threshold interval $[-2,0)$ provides a wide buffer before local modal growth sets in. Reducing the solvent ratio sixfold to $\beta=1/9$ at $\Wi=50$ ($N=257$, $\mathrm{Re}=1$) contracts the nondiffusive frozen-state threshold $-\beta/(1-\beta)$ from $-2$ to $-0.125$, leaving very little margin between $\lmin=0$ and local modal growth. Under Eq.~\eqref{eq:nondim-groups} and Section~\ref{sec:stopping}, lowering $\beta$ by a factor of $6$ at fixed $\mathrm{Sc}=10^5$ and $s_F=1$ simultaneously scales down both $F^*$ and $\kappa^*$ to $1/6$ of their $\beta=2/3$ reference values $F_0^*=4/3$ and $\kappa_0^*=(2/3)10^{-5}$. We therefore compare three unmodified SM-1 parameter scalings in Table~\ref{tab:lowbeta-parameters} and Figure~\ref{fig:lowbeta}: a \emph{weak-forcing} case keeping the raw forcing factor $s_F=1$ ($\mathrm{Sc}=10^5$), a \emph{force-matched} case ($s_F=6$, $\mathrm{Sc}=10^5$, restoring $F^*=F_0^*$), and a \emph{force- and diffusivity-matched} case ($s_F=6$, $\mathrm{Sc}=10^5/6$, restoring both $F^*=F_0^*$ and $\kappa^*=\kappa_0^*$).

\begin{table}[htbp]
\centering\linespread{0.88}\selectfont\scriptsize
\setlength{\tabcolsep}{3pt}
\begin{tabular}{@{}>{\raggedright\arraybackslash}p{\dimexpr0.22\textwidth-2\tabcolsep\relax}>{\raggedright\arraybackslash}p{\dimexpr0.10\textwidth-2\tabcolsep\relax}>{\raggedright\arraybackslash}p{\dimexpr0.11\textwidth-2\tabcolsep\relax}>{\raggedright\arraybackslash}p{\dimexpr0.10\textwidth-2\tabcolsep\relax}>{\raggedright\arraybackslash}p{\dimexpr0.12\textwidth-2\tabcolsep\relax}>{\raggedright\arraybackslash}p{\dimexpr0.11\textwidth-2\tabcolsep\relax}>{\raggedright\arraybackslash}p{\dimexpr0.11\textwidth-2\tabcolsep\relax}>{\raggedright\arraybackslash}p{\dimexpr0.13\textwidth-2\tabcolsep\relax}@{}}
\toprule
Control case ($\beta$, $\mathrm{Sc}$) & $F^*/F_0^*$ & $\kappa^*/\kappa_0^*$ & First $\lmin<0$ & First $\lmin<\ell_{\mathrm{th}}$ & First $U_{\max}>50$ & Coarse exit & Outcome\\
\midrule
Reference ($2/3$, $10^5$)$^{\dagger}$ & $1$ & $1$ & $4.56$ & $5.49$ & -- & $120$ & Reaches $t^*=120$\\
Weak-forcing ($1/9$, $10^5$) & $1/6$ & $1/6$ & $79.27$ & $82.39$ & not reached & $120$ & Reaches $t^*=120$\\
Force-matched ($1/9$, $10^5$) & $1$ & $1/6$ & $2.83$ & $2.98$ & $44.94$ & $45.00$ & Terminated\\
Force- \& $\kappa$-matched ($1/9$, $10^5/6$) & $1$ & $1$ & $2.78$ & $3.06$ & $49.93$ & $50.00$ & Terminated\\
\bottomrule
\end{tabular}
\caption{Parameter scalings and step-resolved event times $t^*$ (in units of $T_c$) for unmodified SM-1 low-solvent control cases at $\Wi=50$, $N=257$, and $\mathrm{Re}=1$, relative to $F_0^*=4/3$ and $\kappa_0^*=(2/3)10^{-5}$ ($\ell_{\mathrm{th}}=-\beta/(1-\beta)$ equals $-2$ at $\beta=2/3$ and $-0.125$ at $\beta=1/9$; TRT relaxation rates follow each physical transport coefficient, and Appendix~\ref{app:revision-checks} documents input matching to the archived $12$-digit decimal reference state). In the two terminating $\beta=1/9$ runs, nonfinite conformations first appear at $t^*=44.97$ and $49.95$. $^{\dagger}$For $\beta=2/3$, every-step monitoring covers the initial prefix $t^*\in[0,8.00]$, while $t^*=120$ continuation without crossing $U_{\max}>50$ (denoted ``--'') is recorded from the archived $\Delta t^*\approx 1$ run.}
\label{tab:lowbeta-parameters}
\label{tab:lowbeta-events}
\end{table}

Under weak forcing ($s_F=1$), unmodified SM-1 reaches $t^*=120$ ($\min\lmin=-3.99$, $\phi_-=0.67\%$, $\max_{\bm x}|\bm u|/U_c=0.0455$; Table~\ref{tab:lowbeta-events} and Figure~\ref{fig:lowbeta}), as sixfold weaker forcing keeps conformation excursions mild while delaying initial SPD loss until $t^*=79.27$ and threshold crossing ($-0.125$) until $t^*=82.39$, leaving only $37.6T_c$ of post-threshold evolution sampled before the $t^*=120$ cutoff. Restoring baseline forcing ($s_F=6$) advances initial departure from $\mathcal K_h$ to $t^*\simeq 2.8$ and reverses the finite-horizon outcome: the force-matched run crosses $U_{\max}>50$ at $t^*=44.94$ ($\mathrm{Ma}\simeq 0.521$, coarse exit $45.00$), and restoring $\kappa^*=\kappa_0^*$ as well delays $U_{\max}>50$ by only $4.98T_c$ to $t^*=49.93$ ($\mathrm{Ma}\simeq 0.509$, coarse exit $50.00$) without reaching $t^*=120$. Both baseline-forcing runs remain finite at $t^*=40.0$ ($\min\lmin=-64.1$ and $-43.0$; peak speeds $0.288U_c$ and $0.255U_c$) and endure $42.0T_c$ and $46.9T_c$ between threshold crossing and $U_{\max}>50$, longer than the $37.6T_c$ post-threshold window sampled under weak forcing.

Equations~\eqref{eq:rate} and~\eqref{eq:cutoff} explain why matched baseline forcing (with fluid and conformation TRT relaxation rates updated alongside $\beta$ and $\kappa^*$) reverses uncontrolled outside-cone survival of SM-1. Lowering $\beta$ from $2/3$ to $1/9$ not only shrinks the sub-threshold buffer sixteenfold to $[-0.125,0)$, but also amplifies the polymer-to-solvent slope $(1-\beta)/\beta=\mu_p/\mu_s$ in Eq.~\eqref{eq:rate} sixteenfold from $1/2$ to $8$. Consequently, at $t^*=40.0$, $\min\lmin=-64.1$ and $-43.0$ already generate frozen plateau rates $\sigma_{\max}T_c=\chi/\Wi\simeq 10.2$ and $6.86$ (versus $1.49$ at the $t^*=120$ endpoint $\min\lmin=-150.5$ for $\beta=2/3$) while expanding the diffusive cutoff $k_c$ in Eq.~\eqref{eq:cutoff} more than twofold even at matched $\kappa^*$.

\FloatBarrier

\subsubsection{Finite-horizon stabilization via shifted-cone projection and smooth spectral restoring}
\label{sec:shifted-cone}

\begin{figure}[htbp]
\centering
\includegraphics[width=0.99\textwidth]{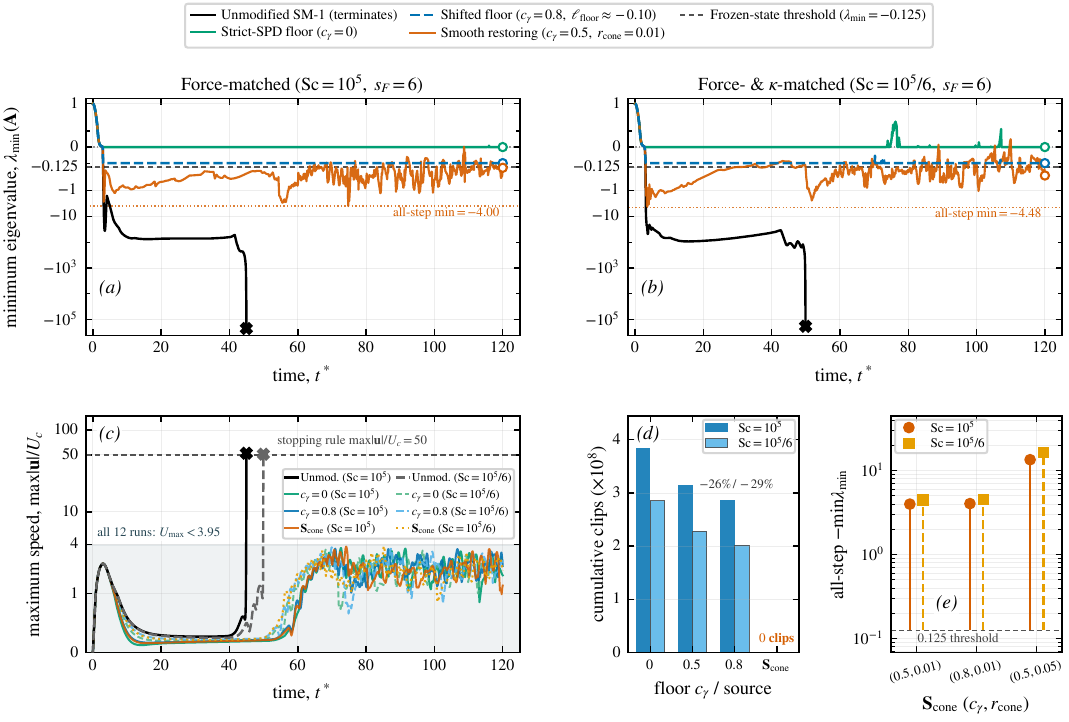}
\caption{Finite-horizon stabilization of SM-1 runs under baseline forcing at low solvent viscosity ($\beta=1/9$, $\Wi=50$, $N=257$, $\mathrm{Re}=1$) by shifted-cone projection $\Pi_{-\gamma+\varepsilon}$ and the smooth spectral-restoring source $\bm S_{\mathrm{cone}}(\A)$ (summary metrics in Table~\ref{tab:shifted-cone}). Panels (a,b) show minimum conformation eigenvalue histories $\lmin(\A)(t^*)$ in the force-matched ($\mathrm{Sc}=10^5$, $s_F=6$) and force- and diffusivity-matched ($\mathrm{Sc}=10^5/6$, $s_F=6$) cases (symmetric logarithmic scale, linear within $|\lmin|\leq 0.125$), comparing the unmodified SM-1 baseline ($\times$ at first velocity-threshold crossing: $t^*=44.94$ and $49.93$) against $c_\gamma=0$, $c_\gamma=0.8$, and $(c_\gamma,r_{\mathrm{cone}})=(0.5,0.01)$, where horizontal dotted lines mark the all-step minimum of each $(0.5,0.01)$ restoring run attained between $0.1T_c$ samples. Panel (c) compares dimensionless peak speeds $\max_{\bm x}|\bm u|/U_c$ (symmetric logarithmic scale) across the unmodified runs, $c_\gamma=0$, $c_\gamma=0.8$, and $\bm S_{\mathrm{cone}}(\A)$ at $(0.5,0.01)$, where unmodified runs cross the stopping threshold $50$ while all twelve stabilized runs (Table~\ref{tab:shifted-cone}) remain bounded by $\max_{t^*} U_{\max}<3.95$ within the same unsteady band; panels (d) and (e) compare cumulative nodal hard-projection counts across $c_\gamma\in\{0,0.5,0.8\}$ and $\bm S_{\mathrm{cone}}$ and all-step excursion depth $-\min\lmin$ across smooth-restoring pairs $(c_\gamma,r_{\mathrm{cone}})$ relative to $\beta/(1-\beta)=0.125$.}
\label{fig:shifted-cone}
\end{figure}

Although unmodified SM-1 terminates in both baseline-forcing cases at $\beta=1/9$, applying either $\Pi_{-\gamma+\varepsilon}$ ($c_\gamma\in\{0,0.5,0.8\}$) or $\bm S_{\mathrm{cone}}(\A)$ carries all twelve low-solvent runs in Table~\ref{tab:shifted-cone} and Figure~\ref{fig:shifted-cone} (two strict-SPD controls at $c_\gamma=0$ and ten runs admitting $\lmin(\A)<0$ at $c_\gamma\in\{0.5,0.8\}$) to $t^*=120$. Peak dimensionless speeds remain bounded within $\max_{t^*} U_{\max}\simeq 3.39$--$3.94$ ($\max_{t^*} U_{\max}<3.95$, $\mathrm{Ma}_{\max}<0.040$; quadruple-precision near-boundary checks and cross-build reproducibility are reported in Appendix~\ref{app:revision-checks}). Under the strict-SPD floor ($c_\gamma=0$), every nodal tensor stays strictly positive definite at every step in both baseline-forcing cases (Table~\ref{tab:shifted-cone} and Figure~\ref{fig:shifted-cone}a,b). At $c_\gamma=0.5$ ($\ell_{\mathrm{floor}}\approx -0.0625$) and $c_\gamma=0.8$ ($\ell_{\mathrm{floor}}\approx -0.10$), the shifted floor leaves the Oldroyd--B update unaltered inside $\mathcal K_{-\gamma+\varepsilon}$ while satisfying $\|\Pi_{-\gamma+\varepsilon}(\A)-\A\|_F\leq\|\Pi_{\varepsilon}(\A)-\A\|_F$ pointwise: it admits negative eigenvalues over peak node fractions of $21.4\%$--$33.3\%$, prevents any excursion past $-0.125$, and reduces cumulative hard projections by $18.3\%$--$29.4\%$ relative to $c_\gamma=0$ (Figure~\ref{fig:shifted-cone}d). These shifted-floor SM-1 runs show that crossing $\lmin(\A)=0$ does not by itself cause low-solvent SM-1 breakdown: keeping $\lmin(\A)$ above $-\beta/(1-\beta)=-0.125$ is sufficient to regain bounded continuation to $t^*=120$ in both baseline-forcing configurations, complementing the $\beta=2/3$ FD-TC runs in Table~\ref{tab:clip-timeline} where keeping stress-supplying eigenvalues strictly positive or above $-1$ (well above $-2$) up to the velocity event still fails under FD-TC force-divergence coupling.

\begin{table}[htbp]
\centering\linespread{0.92}\selectfont\scriptsize
\setlength{\tabcolsep}{3pt}
\renewcommand{\arraystretch}{0.95}
\begin{tabular}{@{}>{\raggedright\arraybackslash}p{\dimexpr0.225\textwidth-2\tabcolsep\relax}>{\raggedright\arraybackslash}p{\dimexpr0.145\textwidth-2\tabcolsep\relax}>{\raggedright\arraybackslash}p{\dimexpr0.12\textwidth-2\tabcolsep\relax}>{\raggedright\arraybackslash}p{\dimexpr0.145\textwidth-2\tabcolsep\relax}>{\raggedright\arraybackslash}p{\dimexpr0.145\textwidth-2\tabcolsep\relax}>{\raggedright\arraybackslash}p{\dimexpr0.13\textwidth-2\tabcolsep\relax}>{\raggedright\arraybackslash}p{\dimexpr0.09\textwidth-2\tabcolsep\relax}@{}}
\toprule
Low-$\beta$ case / operator & Outcome ($t^*\leq 120$) & All-step $\min\lmin$ & Peak / final $\phi_-$ & Peak / late $\phi_{\mathrm{sup}}$ & Cumulative clips & Peak $U_{\max}$\\
\midrule
\multicolumn{7}{@{}>{\raggedright\arraybackslash}p{\textwidth}@{}}{\emph{Force-matched case ($\beta=1/9$, $\mathrm{Sc}=10^5$, $s_F=6$; unmodified baseline exits at $t^*=45.00$)}}\\
Strict-SPD floor ($c_\gamma=0$) & Reaches $t^*=120$ & $+1.00\times10^{-8}$ & $0\%$ / $0\%$ & $0\%$ / $0\%$ & $3.85\times10^8$ & $3.62$\\
Shifted floor ($c_\gamma=0.5$) & Reaches $t^*=120$ & $-0.0625$ & $31.4\%$ / $0.11\%$ & $0\%$ / $0\%$ & $3.14\times10^8$ & $3.79$\\
Shifted floor ($c_\gamma=0.8$) & Reaches $t^*=120$ & $-0.1000$ & $33.3\%$ / $1.23\%$ & $0\%$ / $0\%$ & $2.86\times10^8$ & $3.55$\\
Restoring ($c_\gamma=0.5$, $r_{\mathrm{cone}}=0.01$) & Reaches $t^*=120$ & $-4.00$ & $32.3\%$ / $0.98\%$ & $2.98\%$ / $0.033\%$ & $0$ & $3.77$\\
Restoring ($c_\gamma=0.8$, $r_{\mathrm{cone}}=0.01$) & Reaches $t^*=120$ & $-4.05$ & $33.8\%$ / $1.47\%$ & $7.97\%$ / $0.130\%$ & $0$ & $3.55$\\
Restoring ($c_\gamma=0.5$, $r_{\mathrm{cone}}=0.05$) & Reaches $t^*=120$ & $-13.5$ & $34.6\%$ / $0.90\%$ & $16.8\%$ / $0.323\%$ & $0$ & $3.94$\\
\midrule
\multicolumn{7}{@{}>{\raggedright\arraybackslash}p{\textwidth}@{}}{\emph{Force- and diffusivity-matched case ($\beta=1/9$, $\mathrm{Sc}=10^5/6$, $s_F=6$; unmodified baseline exits at $t^*=50.00$)}}\\
Strict-SPD floor ($c_\gamma=0$) & Reaches $t^*=120$ & $+1.00\times10^{-8}$ & $0\%$ / $0\%$ & $0\%$ / $0\%$ & $2.85\times10^8$ & $3.45$\\
Shifted floor ($c_\gamma=0.5$) & Reaches $t^*=120$ & $-0.0625$ & $21.4\%$ / $0.85\%$ & $0\%$ / $0\%$ & $2.28\times10^8$ & $3.73$\\
Shifted floor ($c_\gamma=0.8$) & Reaches $t^*=120$ & $-0.1000$ & $24.7\%$ / $0.26\%$ & $0\%$ / $0\%$ & $2.02\times10^8$ & $3.69$\\
Restoring ($c_\gamma=0.5$, $r_{\mathrm{cone}}=0.01$) & Reaches $t^*=120$ & $-4.48$ & $21.8\%$ / $1.00\%$ & $1.42\%$ / $0.020\%$ & $0$ & $3.83$\\
Restoring ($c_\gamma=0.8$, $r_{\mathrm{cone}}=0.01$) & Reaches $t^*=120$ & $-4.55$ & $24.8\%$ / $0.71\%$ & $2.26\%$ / $0.061\%$ & $0$ & $3.52$\\
Restoring ($c_\gamma=0.5$, $r_{\mathrm{cone}}=0.05$) & Reaches $t^*=120$ & $-16.6$ & $23.8\%$ / $1.21\%$ & $3.95\%$ / $0.142\%$ & $0$ & $3.39$\\
\bottomrule
\end{tabular}
\caption{Every-step diagnostics over steps $0$--$850150$ ($850151$ audited updates to $t^*\simeq 120$) for strict-SPD ($c_\gamma=0$), shifted-cone ($c_\gamma\in\{0.5,0.8\}$), and smooth spectral-restoring runs under SM-1 at $\beta=1/9$, $N=257$, $\Wi=50$, and $\mathrm{Re}=1$. Here $\phi_-$ and $\phi_{\mathrm{sup}}$ denote domain node fractions with $\lmin<0$ and $\lmin<-0.125$, ``late $\phi_{\mathrm{sup}}$'' is the all-step mean over $t^*\geq 60$ ($425076$ updates), and $U_{\max}=\max_{\bm x}|\bm u|/U_c$. All twelve runs fail the tail-stationarity test of Section~\ref{sec:stopping}.}
\label{tab:shifted-cone}
\end{table}

By contrast, the smooth source $\bm S_{\mathrm{cone}}(\A)$ eliminates hard clipping ($0$ projections across all six restoring runs) and restores bounded continuation to $t^*=120$ via rapid state pull-back alongside the linear transverse damping $-g_{\mathrm{cone}}^{[1]}(a_1,a_2)\geq 0$ of Remark~\ref{rem:modified-pde} (Figure~\ref{fig:shifted-cone}a,b,e). Because $\tau_{\mathrm{cone}}\approx 70.85\Delta t$ ($r_{\mathrm{cone}}=0.01$) and $354.23\Delta t$ ($r_{\mathrm{cone}}=0.05$), recurring dips past $-0.125$ appear on $64.7\%$--$99.4\%$ of late-time steps ($t^*\in[60,120]$), but remain confined to small mean supercritical areas $\phi_{\mathrm{sup}}$ of $0.020\%$--$0.323\%$ ($\lmin<-0.125$, Table~\ref{tab:shifted-cone}), with all-step minima reaching $-4.00$ and $-4.48$ at $(0.5,0.01)$ and deepening to $-13.5$ and $-16.6$ at $(0.5,0.05)$. Even at those transient extrema, restoring pull-back outpaces local frozen growth ($\sigma_{\max}\tau_{\mathrm{cone}}\simeq 6\times 10^{-3}$--$7\times 10^{-3}\ll 1$ at $(0.5,0.01)$ and $\leq 0.14<1$ across all six restoring runs).

Late-time flow observables over $t^*\in[60,120]$ further contrast operator agreement at moderate and low solvent viscosity. At $\beta=2/3$ (Table~\ref{tab:matrix}), $\Pi_{-1+\varepsilon}$ and $\bm S_{\mathrm{cone}}(\A)$ agree within $0.1\%$ in mean $U_{\max}$ ($0.2235$ versus $0.2234$) and about $0.3\%$ in mean $\Wi_x$ ($0.612$ versus $0.610$). At $\beta=1/9$, all twelve stabilized runs remain strongly unsteady (failing the tail-stationarity test of Section~\ref{sec:stopping}) within the same speed band as strict-SPD $c_\gamma=0$ (Figure~\ref{fig:shifted-cone}c): mean $U_{\max}$ spans $2.00$--$2.24$ ($2.13\pm 0.59$ at $c_\gamma=0$) in the force-matched set and $1.91$--$2.14$ ($1.91\pm 0.50$ at $c_\gamma=0$) in the force- and diffusivity-matched set. The finite-window mean $\Wi_x$ spans $35$--$71$ across operators, comparable to the realization spread ($66.3$ versus $49.4$) between two force-matched strict-SPD ($c_\gamma=0$) runs on the same binary whose forcing inputs differ relatively by $5\times 10^{-13}$ (Appendix~\ref{app:revision-checks}). With $R_\Delta=\Delta x/\ell_d\simeq 3.3$ and $1.3$ on $N=257$, these twelve runs represent bounded, regularized continuations of an under-resolved unsteady flow rather than grid-converged Oldroyd--B solutions.

\FloatBarrier

%==============================================================================
\section{Discussion}
\label{sec:discussion}

\subsection{Separating the three diagnostic questions across solvent and diffusion limits}
\label{sec:diagnosis}

Together, the analytical and numerical results separate three distinct questions: whether the continuum PDE preserves $\mathcal K^{\circ}$ (answered by Proposition~\ref{prop:continuum-cone}), whether an indefinite frozen state amplifies local transverse perturbations (answered by Proposition~\ref{prop:threshold} and Eq.~\eqref{eq:cutoff}), and whether a coupled discrete solver survives over a prescribed horizon (answered by the four-roll-mill experiments of Section~\ref{sec:causal}). Keeping these three levels distinct explains why identical eigenvalue signs accompany opposite computational outcomes across different discretizations and solvent ratios.

As $\beta\to 0$ along the nondiffusive frozen branch, the threshold $-\beta/(1-\beta)$ vanishes while the plateau rate $\sigma_{\max}$ in Eq.~\eqref{eq:rate} diverges, recovering the unbounded $\sigma\propto k$ Maxwell instability of Corollary~\ref{cor:ucm} at $\beta=0$ where any negative eigenvalue triggers arbitrarily rapid short-wave growth. Fixed positive diffusivity $\kappa>0$ as $\beta\to 0$ caps high-wavenumber growth by Corollary~\ref{cor:ucm}, whereas fixed $\mathrm{Sc}$ scales $\kappa^*=\beta/(\mathrm{Re}\,\mathrm{Sc})$ proportionally to $\beta$, a parameter coupling isolated directly by the matched-forcing and matched-diffusivity controls in Section~\ref{sec:lowbeta}.

\subsection{Relation to positive \mbox{(semi-)}def\-i\-nite reformulations and statistical evidence}
\label{sec:related-mechanisms}

Log-conformation, square-root, and Cholesky factorizations enforce positive \mbox{(semi-)}def\-i\-nite\-ness by construction and exclude indefinite states \cite{VC2003,FK2004,FK2005,Balci2011}, while logarithmic free-energy and entropy functionals involving $\operatorname{tr}(\A-\log\A-\bm I)$ require strictly positive eigenvalues to remain well-defined \cite{HL2007}. Our finding that strict-SPD projection fails to prevent positive-stretch blowup under FD-TC aligns directly with the observation of Hulsen, Spanjaards, and Anderson \cite{HSA2021} that positive \mbox{(semi-)}def\-i\-nite reformulations still exhibit markedly different breakdown limits. Relative to the mechanism of Fattal and Kupferman \cite{FK2004,FK2005} examined in \cite{HSA2021}, where under-resolved, nearly exponential stress profiles in high-residence-time regions trigger instability, our results distinguish in-cone breakdown from modal growth after constraint relaxation across steep stress layers. When steep positive stress layers interact either with unbalanced constitutive convection--deformation discretizations in positivity-preserving formulations \cite{HSA2021} or with force-divergence momentum coupling in strict-SPD-projected FD-TC, runaway develops directly inside the SPD cone $\mathcal K^{\circ}$ while every conformation tensor supplying polymer stress remains strictly positive definite. Under derivative-free local stress-moment coupling (SM-1), under-resolved discrete transport across those layers instead drives the state out of the discrete realizability cone $\mathcal K_h$ into $\mathcal K_h^c$ via constraint relaxation across the coupled collision--streaming update (Section~\ref{sec:first-cross}); once the discrete state leaves $\mathcal K_h$, the frozen-state threshold $-\beta/(1-\beta)$ and coupling slope $(1-\beta)/\beta$ account for why indefinite excursions remain bounded at $\beta=2/3$, lead to breakdown under matched forcing at $\beta=1/9$, and return to bounded continuation once eigenvalues are kept above $-\beta/(1-\beta)$ without resolving the steep positive profile itself, even though the local frozen-state rate alone does not predict how deep localized super-threshold excursions can grow before global breakdown.

\label{sec:capocci}Capocci, Linkmann, and Morozov \cite{NoNeed} simulated three-dimensional simplified Phan-Thien--Tanner (sPTT) channel flow with polymer-stress diffusion at $\beta=0.9$ and $\Wi=150$ (in their notation, conformation tensor $\bm c$, laminar centerline speed $U_0$, diffusivity $D$ with $\kappa_C=D/(dU_0)=5\times 10^{-5}$, extensibility parameter $\epsilon=5\times 10^{-5}$, and horizontal period $10d$ in half-width units $d$). Their simulations exhibit two resolution thresholds that are consistent with the present framework: resolutions below \texttt{A1} (not tabulated) blow up in finite time, intermediate grids \texttt{A1}--\texttt{A5} remain stable despite $\min(\operatorname{tr}\bm c-3)\simeq -1800$ ($\lmin(\bm c)\lesssim -599$ from $\lmin(\bm c)\leq\operatorname{tr}\bm c/3$, past $-9$), and \texttt{A6} is reported to remain strictly SPD throughout. First, in Eq.~\eqref{eq:rate}, the Oldroyd--B coupling slope $(1-\beta)/(\beta\Wi)=1/1350$ is $13.5$ and $216$ times smaller than at $(\beta,\Wi)=(2/3,50)$ ($1/100$) and $(1/9,50)$ ($8/50$): at the equal-eigenvalue reference $\lmin=-599$, Eq.~\eqref{eq:rate} gives $\sigma_{\max}d/U_0\simeq 0.44$ with cutoff $2\pi/k_c\simeq 0.067d$ in Eq.~\eqref{eq:cutoff}, while the sPTT factor $f_0=1+\epsilon\operatorname{tr}(\bm c-\bm I)\simeq 0.91>0$ stays positive (restorative) yet below unity (shifting the eigenvector-aligned transverse threshold from $-9$ to $-9f_0\simeq -8.19$, with $>1$ damping in $\operatorname{tr}\bm c\gg 3$ sheets). Because the trace yields only the one-sided bound $\lmin\lesssim -599$ ($\sigma_{\max}d/U_0\gtrsim 0.44$, $2\pi/k_c\lesssim 0.067d$; exact modal ordering requires full spectra), local rates alone do not ensure stability. Second, $\ell_d=\sqrt{D\lambda}=d\sqrt{\kappa_C\Wi}\simeq 0.087d$ provides a reference diffusive scale across their tabulated resolutions: midplane Chebyshev spacing satisfies $\Delta y_{\mathrm{mid}}/\ell_d\leq 0.29$ throughout, whereas nominal horizontal Fourier spacing $\max(\Delta x,\Delta z)/\ell_d$ sits near unity ($\simeq 0.90$) at their coarsest stable resolution \texttt{A1} ($N_x=N_z=128$; a twofold coarser horizontal mode count of $64$ would give $\simeq 1.80>1$) and drops to $\simeq 0.23$ on \texttt{A6} ($N_x=N_z=512$), mirroring Table~\ref{tab:field-sensitivity}, although doubling streamwise modes alone between \texttt{A3} and \texttt{A5} at identical $\max(\Delta x,\Delta z)/\ell_d\simeq 0.45$ lowers violation incidence by three orders of magnitude, showing that anisotropic streak resolution is not captured by $\max(\Delta x,\Delta z)/\ell_d$ alone. Third, in \texttt{A1}--\texttt{A5}, $\operatorname{tr}\bm c<3$ violations (a midplane trace metric distinct from $\phi_-$) remain confined to stagnation-zone pockets with per-sample probability $\mathcal P(\operatorname{tr}\bm c<3)\leq 6.6\times 10^{-5}$ and cumulative midplane event ratio at most $7.2\times 10^{-4}$ ($0.072\%$), leaving $>99.9\%$ of midplane samples in $\operatorname{tr}\bm c>3$ and preserving velocity and velocity-gradient statistics.

\subsection{Implications for algorithm design and method evaluation}
\label{sec:practice}\label{sec:implications}

For algorithm design and benchmarking, hard projection, spatial filtering, spectral-restoring sources, and variable transformations each alter both realizability enforcement and momentum forcing. Within polynomial closures such as Oldroyd--B, the two threshold-anchored operators serve distinct roles: $\Pi_{-\gamma+\varepsilon}$ at $c_\gamma\in(0,1)$ provides a minimum-Frobenius-perturbation diagnostic control that tests confinement above $-\beta/(1-\beta)$ separately from $\lmin(\A)>0$ without modifying the interior PDE, whereas $\bm S_{\mathrm{cone}}(\A)$ provides a projection-free $C^\infty$ regularization that trades outside-cone PDE fidelity for smooth, kink-free continuation. Because polynomial stress evaluation remains well-defined across zero eigenvalues, $\bm S_{\mathrm{cone}}(\A)$ is defined for both $c_\gamma=0$ and $c_\gamma\in(0,1)$ (tested here at $c_\gamma\in\{0.5,0.8\}$), extending smooth spectral control into the sub-threshold nonpositive regime where the principal real symmetric matrix logarithm $\log\A$ and logarithmic free-energy functionals involving $\operatorname{tr}(\A-\log\A-\bm I)$ are undefined. Benchmarks of high-Weissenberg-number schemes should therefore test realizability control jointly with discrete stress--momentum coupling at matched physical and grid parameters, judging candidate methods by accuracy, statistical stationarity, and grid convergence rather than by eigenvalue sign or finite-horizon survival alone; the low-solvent interventions reported here establish finite-horizon boundedness, whereas grid convergence in that unsteady regime remains to be tested.

%==============================================================================
\section{Conclusions}
\label{sec:concl}

Negative conformation eigenvalues in polynomial-closure simulations starting from $\A=\bm I$ represent discrete constraint relaxation relative to corresponding periodic diffusive Oldroyd--B solutions, which remain strictly inside $\mathcal K^{\circ}$. Locally about a uniform frozen tensor for $0<\beta<1$, nondiffusive transverse eigenmodes grow only when $\lmin(\A)<-\beta/(1-\beta)$, with growth rates approaching a finite positive short-wave plateau, whereas positive stress diffusion restricts growing modes to $k<k_c$. Independently, two-dimensional constitutive relaxation contributes a positive determinant-restoring rate $H_{\mathrm{rel}}>0$ wherever $\det\A<0$ and $\operatorname{tr}\A>0$.

In coupled four-roll-mill computations at high Weissenberg number, finite-horizon survival depends jointly on discrete stress--momentum coupling and on bounding super-threshold spectral excursions; strict SPD preservation is neither necessary nor sufficient. At moderate solvent viscosity ($\beta=2/3$, $\Wi=50$), centered-difference time-consistent force-divergence coupling (FD-TC) crosses the velocity stopping threshold even when every conformation eigenvalue supplying polymer stress is held strictly positive or confined above $-1$ (well above $-2$) through that crossing, whereas derivative-free local stress-moment schemes and the Guo force-divergence scheme maintain bounded continuation to $t^*=120$ despite deep outside-cone excursions. At low solvent viscosity ($\beta=1/9$), under matched baseline forcing where unmodified SM-1 terminates, Frobenius-orthogonal strict-SPD and shifted-cone projections $\Pi_{-\gamma+\varepsilon}$ ($c_\gamma\in\{0,0.5,0.8\}$) and the smooth softplus spectral-restoring source $\bm S_{\mathrm{cone}}(\A)$ sustain bounded unsteady evolution to $t^*=120$ across all twelve stabilized SM-1 runs (two strict-SPD controls and ten runs admitting $\lmin(\A)<0$). Viscoelastic discretizations should therefore be designed and evaluated as coupled momentum--constitutive systems. Extending this threshold-anchored framework to wall-bounded geometries, three-dimensional flows, and finitely extensible closures provides a natural direction for future study.

\section*{Acknowledgments}
This work was financially supported by the National Natural Science Foundation of China
(Grant No.\ 12101527), the Natural Science Foundation of Hunan Province (Grant No.\
2026JJ60005), the Science and Technology Innovation Program of Hunan Province (Program
No.\ 2026RC3172), and the 111 Project (Grant No.\ D23017). Computational resources were
provided by the High Performance Computing Platform of Xiangtan University. The funding
sources had no involvement in the study design, in the collection, analysis, and
interpretation of data, in the writing of the report, or in the decision to submit the
article for publication.

\section*{Declaration of competing interest}
The authors declare that they have no known competing financial interests or
personal relationships that could have appeared to influence the work reported
in this paper.

\section*{Declaration of generative AI and AI-assisted technologies in the manuscript preparation process}
During the preparation of this work the authors used Claude (Anthropic) to
assist with drafting, language editing, translation, and internal consistency
checks. After using this tool, the authors reviewed and
edited the content as needed and take full responsibility for the content of
the published article.

\section*{Data availability}
The baseline data and scripts from the initial submission are openly available in the Zenodo repository \cite{YuDataset2026} (version~1.0.0). The supplementary archive for this revision contains the solver sources, simulation histories, step-resolved event logs, and verification scripts for the strict-SPD reruns, smooth Kolmogorov verification, and low-solvent spectral-control studies (\href{https://doi.org/10.5281/zenodo.24356270}{doi:10.5281/zenodo.24356270}).

\appendix
\labelformat{section}{\Alph{section}}
\labelformat{subsection}{\Alph{section}.\arabic{subsection}}
\section{Derivation of the transverse dispersion relation}
\label{app:disp}

\subsection{Transverse reduction of the linearized equations}

Along the spatially uniform unforced background of Eq.~\eqref{eq:relaxing-background}, fix $\A_0=\overline{\A}(t_0)$ at a chosen instant $t_0$, where $\A_0$ is symmetric but need not lie in $\mathcal K$. For a plane-wave perturbation $(\delta\bm u,\delta\!\A,\delta p)\propto e^{i\bm k\cdot\bm x+\sigma t}$ with $k=|\bm k|>0$ and $\khat=\bm k/k$, incompressibility confines $\delta\bm u=\sum_j v_j\hat{\bm t}_j$ to an orthonormal basis $\{\hat{\bm t}_j\}$ perpendicular to $\khat$. Since $(\nabla\delta\bm u)_{mn}=ik\,\hat t_{j,m}\hat k_n v_j$ (summing over repeated transverse indices $j$), the linearized upper-convected stretching tensor reads
\begin{equation}
(\nabla\delta\bm u)\A_0+\A_0(\nabla\delta\bm u)^{\!\top}
=ikv_j\big(\hat{\bm t}_j\otimes\A_0\khat
+\A_0\khat\otimes\hat{\bm t}_j\big).
\end{equation}
Projecting $\delta\!\A$ onto the mixed transverse--longitudinal components $\zeta_i=\hat{\bm t}_i^{\!\top}\delta\!\A\,\khat$ and using $\hat{\bm t}_j^{\!\top}\khat=0$ gives
\begin{equation}
\Big(\sigma+\frac{1}{\lambda}+\kappa k^2\Big)\zeta_i
=ik\,v_j\Big[\delta_{ij}(\khat^{\!\top}\!\A_0\khat)
+(\hat{\bm t}_i^{\!\top}\!\A_0\khat)
(\hat{\bm t}_j^{\!\top}\khat)\Big]
=ik\,v_i(\khat^{\!\top}\!\A_0\khat).
\label{eq:appA-const}
\end{equation}
Adding a uniform background velocity $\bm u_0$ contributes only the purely imaginary advection shift $-i\bm k\cdot\bm u_0$ to $\sigma$ without altering $\operatorname{Re}\sigma$. Since $\delta\bm\tau=(\mu_p/\lambda)\delta\!\A$ and $\delta p$ drops out under transverse projection, $\hat{\bm t}_i^{\!\top}(i\bm k\cdot\delta\bm\tau)=(\mu_p/\lambda)ik\zeta_i$ and the transverse momentum balance becomes
\begin{equation}
(\rho\sigma+\mu_s k^2)v_i=\frac{\mu_p}{\lambda}ik\zeta_i .
\label{eq:appA-mom}
\end{equation}
Equations~\eqref{eq:appA-const} and~\eqref{eq:appA-mom} decouple all $(d{-}1)$ transverse polarizations into identical $2\times 2$ scalar blocks, so two- and three-dimensional transverse modes obey the same dispersion law.

\subsection{Dispersion relation and exact threshold}

Eliminating $(v_i,\zeta_i)$ between Eqs.~\eqref{eq:appA-const} and~\eqref{eq:appA-mom} yields Eq.~\eqref{eq:disp}. Setting $a=\khat^{\!\top}\!\A_0\khat$ casts Eq.~\eqref{eq:disp} for $\rho>0$ into the quadratic $\rho\sigma^2+b\sigma+c=0$ with
\begin{equation*}
b=\mu_s k^2+\rho\Big(\frac{1}{\lambda}+\kappa k^2\Big)>0,
\qquad
c=\mu_s k^2\Big(\frac{1}{\lambda}+\kappa k^2\Big)
+\frac{\mu_p}{\lambda}k^2 a .
\end{equation*}
Because $b>0$, a root with $\operatorname{Re}\sigma>0$ exists if and only if $c<0$ (while $c>0$ damps both roots and $c=0$ yields a neutral root); setting $\rho=0$ with $\mu_s>0$ gives the identical sign condition from the linear Stokes relation. For every $k>0$, modal growth therefore occurs if and only if
\begin{equation}
\mu_s\big(1+\kappa\lambda k^2\big)
<\mu_p\big({-}\khat^{\!\top}\!\A_0\khat\big).
\label{eq:appA-thresh}
\end{equation}
Minimizing $a$ over $|\khat|=1$ at $\kappa=0$ establishes Proposition~\ref{prop:threshold}, and solving Eq.~\eqref{eq:appA-thresh} for $k^2$ at $\kappa>0$ gives Eq.~\eqref{eq:cutoff}.

\subsection{Limit checks}

Setting $\rho=0$ with $\mu_s>0$ in Eq.~\eqref{eq:disp} gives the inertialess diffusive rate
\begin{equation*}
\sigma=\frac{1}{\lambda}
\left[\frac{\mu_p}{\mu_s}(-a)-1\right]-\kappa k^2,
\end{equation*}
which reduces at $\kappa=0$ to the wavenumber-independent plateau $\sigma_\infty(\khat)$ in Eq.~\eqref{eq:rate-directional}. At $\mu_s=\kappa=0$, the squared elastic shear-wave speed $c_{\mathrm e}^2=(\mu_p/\rho\lambda)a$ \cite{Rutkevich1970,JRS1985} turns negative when $a<0$, yielding Corollary~\ref{cor:ucm}, whereas any $\A_0\in\mathcal K^{\circ}$ satisfies $a>0$ for all $\khat$, ensuring $c>0$ and damped elastic shear waves with $c_{\mathrm e}^2>0$.

\section{Determinant identities in two and three dimensions and reconstructed closure residual}
\label{app:det}

\subsection{Adjugate form of the determinant derivative}

For any square matrix $\A$, the polynomial adjugate identities
\begin{equation*}
\A\,\operatorname{adj}\A=\operatorname{adj}\A\,\A
=\det\!\A\,\bm I,
\qquad
\mathrm{d}(\det\A)
=\operatorname{tr}(\operatorname{adj}\A\;\mathrm{d}\A)
\end{equation*}
hold globally across singular and indefinite states. Writing $\bm L=\nabla\bm u$ and $\dot\A_{\mathrm{stretch}}=\bm L\A+\A\bm L^{\!\top}$, trace cyclicity gives in every spatial dimension $d$
\begin{equation}
\begin{split}
\operatorname{tr}\!\left[\operatorname{adj}\A
(\bm L\A+\A\bm L^{\!\top})\right]
&=\operatorname{tr}(\A\,\operatorname{adj}\A\,\bm L)
+\operatorname{tr}(\operatorname{adj}\A\,\A\,\bm L^{\!\top})\\
&=2(\operatorname{tr}\bm L)\det\A ,
\end{split}
\end{equation}
consistent with $\A=\bm F\A_0\bm F^{\!\top}$ and $\mathrm{d}(\det\bm F)/\mathrm{d}t=(\operatorname{tr}\bm L)\det\bm F$ for deformation gradient $\bm F$. For $\dot\A_{\mathrm{relax}}=-(\A-\bm I)/\lambda$, using $\operatorname{tr}(\operatorname{adj}\A\,\A)=d\det\A$ and $\operatorname{tr}(\operatorname{adj}\A)=I_{d-1}(\A)$ (the $(d{-}1)$-th elementary symmetric invariant) yields the dimension-specific relaxation terms.

\subsection{Two- and three-dimensional determinant evolution}

At $\kappa=0$, Eq.~\eqref{eq:conf} implies (with Eq.~\eqref{eq:appB-3d} recovering the smooth nondiffusive invariant identity of Hulsen \cite[Eq.~(18)]{Hulsen1990} for $d=3$)
\begin{align}
d=2:&\quad
\frac{\mathrm{D}\det\A}{\mathrm{D}t}
=2(\nabla\!\cdot\!\bm u)\det\A
-\frac{2}{\lambda}\det\A+\frac{\operatorname{tr}\A}{\lambda},
\label{eq:appB-2d}\\
d=3:&\quad
\frac{\mathrm{D}\det\A}{\mathrm{D}t}
=2(\nabla\!\cdot\!\bm u)\det\A
-\frac{3}{\lambda}\det\A+\frac{I_2(\A)}{\lambda},
\label{eq:appB-3d}
\end{align}
with $I_2(\A)=\tfrac{1}{2}[(\operatorname{tr}\A)^2-\operatorname{tr}(\A^2)]$. Equation~\eqref{eq:appB-2d} yields the two-dimensional positivity property $H_{\mathrm{rel}}>0$ for $\det\A<0<\operatorname{tr}\A$ in Section~\ref{sec:selfheal}. That property does not carry over to all positive-trace indefinite tensors in three dimensions: for eigenvalues $(2/5,2/5,-7/10)$, one has $\operatorname{tr}\A=1/10>0$ and $\det\A=-14/125<0$, yet $I_2(\A)-3\det\A=-8/125<0$, making the relaxation term in Eq.~\eqref{eq:appB-3d} initially negative.

\subsection{Reconstructed closure residual}

Retaining $\kappa>0$ in two dimensions adds $\kappa\operatorname{tr}(\operatorname{adj}\A\,\Delta\A)$ as in Eq.~\eqref{eq:det2d}. When evaluating Eq.~\eqref{eq:det2d} from archived snapshots using discrete temporal and spatial differentiation stencils, we define the reconstructed closure residual
\begin{equation}
\mathcal R_{\mathrm{cl}}(\bm x,t)\equiv
\frac{\mathrm{D}\det\A}{\mathrm{D}t}
-2(\nabla\!\cdot\!\bm u)\det\A
+\frac{2}{\lambda}\det\A
-\frac{\operatorname{tr}\A}{\lambda}
-\kappa\operatorname{tr}\!\left(\operatorname{adj}\A\,\Delta\A\right),
\end{equation}
which vanishes for an exact smooth solution of Eq.~\eqref{eq:conf} but aggregates finite-snapshot time differencing, spatial derivative truncation, and discrete lattice closure residuals in post-processed data (Table~\ref{tab:budget-mask}).

\begin{table}[htbp]
\centering\linespread{0.92}\selectfont
\small
\begin{tabular}{@{}>{\raggedright\arraybackslash}p{\dimexpr0.61\textwidth-2\tabcolsep\relax}>{\raggedleft\arraybackslash}p{\dimexpr0.13\textwidth-2\tabcolsep\relax}>{\raggedleft\arraybackslash}p{\dimexpr0.13\textwidth-2\tabcolsep\relax}>{\raggedleft\arraybackslash}p{\dimexpr0.13\textwidth-2\tabcolsep\relax}@{}}
\toprule
Reconstructed determinant-balance term ($\det\A<0$ subset mean) & O2 & O4 & Fourier \\
\midrule
Unsteady rate $\partial_t\det\A$ & $-4$ & $-4$ & $-4$ \\
Compressibility contribution $2(\nabla\!\cdot\!\bm u)\det\A$ & $14$ & $3$ & $1$ \\
Advective transport $-\bm u\cdot\nabla\det\A$ & $183$ & $145$ & $-82$ \\
Diffusive contribution $\kappa\operatorname{tr}(\operatorname{adj}\A\,\Delta\A)$ & $165$ & $187$ & $192$ \\
Constitutive-relaxation recovery $H_{\mathrm{rel}}=(\operatorname{tr}\A-2\det\A)/\lambda$ & $671$ & $671$ & $671$ \\
Reconstructed balance residual $\mathcal R_{\mathrm{cl}}$ & $-1037$ & $-1010$ & $-786$ \\
\bottomrule
\end{tabular}
\caption{Subset-averaged determinant-balance terms (in determinant units per $T_c$, rounded to the nearest integer) over the $\det\A<0$ subset of the baseline SM-1 field ($\Wi=20$, $N=257$, $\mathrm{Sc}=10^5$, $t^*=95$, Figure~\ref{fig:budget}) using centered three-snapshot temporal differencing and second-order (O2), fourth-order (O4), or spectral Fourier spatial derivatives. Here $\mathcal R_{\mathrm{cl}}$ is evaluated by algebraic closure ($-\mathcal R_{\mathrm{cl}}=1037$, $1010$, and $786$).}
\label{tab:budget-mask}
\end{table}
\FloatBarrier

\section{Auxiliary discrete operators, sub-step stage diagnostics, and verification protocols}
\label{app:methods}

\subsection{Local-test fitting protocol and diffusive Kolmogorov reference}

Table~\ref{tab:local-protocol} specifies the $19$ pseudospectral runs reported in Figure~\ref{fig:local-validation}. All runs initialize $\bm u$ with a transverse cosine velocity mode $u_x$ of amplitude $10^{-9}$ at $\bm k=(0,k)$ and initialize the sine mode of $A_{xy}$ from the corresponding frozen-state eigenvector of Eq.~\eqref{eq:disp}. Growth rates are obtained by linear least-squares regression of $\ln a_k(t)$ with $a_k(t)=(2/N^2)|\widehat u_x(0,k,t)|$ across post-step samples in $[0.15,0.9]T_{\mathrm{run}}$ and compared against Eq.~\eqref{eq:disp} evaluated on the analytical background $\overline{\A}(t)$ at the window midpoint.

\begin{table}[htbp]
\centering\footnotesize
\setlength{\tabcolsep}{4pt}
\begin{tabular}{@{}lcccccc@{}}
\toprule
Series & $\Wi$ & $\beta$ & $\kappa^*$ & $A_{0yy}$ & $k$ & $T_{\mathrm{run}}/T_c$\\
\midrule
Solvent & $4$ & $2/3$ & $0$ & $-10$ & $1,2,4,8$ & $1$\\
UCM & $4$ & $0$ & $0$ & $-10$ & $1,2,4,8$ & $0.35$\\
Threshold & $16$ & $2/3$ & $0$ & $-4,\dots,-1$ & $4$ & $1.2$\\
Diffusion & $4$ & $2/3$ & $1/9$ & $-10$ & $1,2,3,4,6$ & $1$\\
\bottomrule
\end{tabular}
\caption{Parameter protocol for the $19$ pseudospectral local tests in Figure~\ref{fig:local-validation} ($N=64$, $\mathrm{Re}=1$, $A_{0xx}=5$, $A_{0xy}=0$, $\bm k=(0,k)$). The threshold series scans $A_{0yy}\in\{-1,-1.5,-2,-2.5,-3,-4\}$ at $\Wi=16$ to slow background relaxation near $-2$, the UCM series uses a shorter window $T_{\mathrm{run}}/T_c=0.35$ so rapid high-$k$ growth stays in the linear regime, and the diffusion series has midpoint cutoff $k_c\simeq 2.73$.}
\label{tab:local-protocol}
\end{table}

For the Kolmogorov flow of Section~\ref{sec:verify} driven from rest ($\bm u=\bm 0$, $\A=\bm I$) by body force $\bm F_b=F_k(\sin(k_fy),0)$, the exact laminar ansatz $u_x(y,t)=U(t)\sin(k_fy)$, $A_{yy}=1$, $A_{xy}(y,t)=S(t)\cos(k_fy)$, and $A_{xx}(y,t)=1+C_0(t)+C_2(t)\cos(2k_fy)$ reduces Eqs.~\eqref{eq:mom}--\eqref{eq:conf} to the closed linear--quadratic system $\dot U=-\nu_s k_f^2 U-[\mu_p k_f/(\rho\lambda)]S+F_k/\rho$, $\dot S=k_f U-(\lambda^{-1}+\kappa k_f^2)S$, $\dot C_0=k_f U S-\lambda^{-1}C_0$, and $\dot C_2=k_f U S-(\lambda^{-1}+4\kappa k_f^2)C_2$. As $t\to\infty$ with $\delta=\kappa\lambda k_f^2$, this system relaxes to the steady profile $U_\kappa=F_k/[k_f^2(\mu_s+\mu_p/(1+\delta))]$, $A_{xy,\kappa}=\lambda k_f U_\kappa\cos(k_fy)/(1+\delta)$, and $A_{xx,\kappa}=1+[(\lambda k_f U_\kappa)^2/(1+\delta)][1+\cos(2k_fy)/(1+4\delta)]$; broader nonlinear effects of polymer-stress diffusion are analyzed in \cite{GV2019}.

\subsection{Sub-step ordering and auxiliary state-modification operators}
\label{app:intervention}

Within step $n\to n+1$ ($\Delta t$), conformation collision and streaming first advance $g_{ai}^n\to g_{ai}^{n+1}$ from $(\rho^n,\bm u^n,\nabla\bm u^n,\A^n)$, including the body-force flux population $\Delta t(1-\Omega_{p1}/2)(w_i/c_s^2)(A_a^n/\rho^n)(\bm c_i\cdot\bm F_b)$ alongside the constitutive source stencil below. Macroscopic reconstruction $A_a^{n+1}=\sum_i g_{ai}^{n+1}$, optional state filtering or eigenvalue projection via Eqs.~\eqref{eq:shifted-proj}--\eqref{eq:pop-writeback}, evaluation of $\bm\sigma_p^{n+1}$ via Eq.~\eqref{eq:discrete-tau} (and $\bm F_p^{n+1}=\nabla\!\cdot\!\bm\sigma_p^{n+1}$ for FD schemes), and non-intrusive every-step conformation monitoring then precede the hydrodynamic collision--streaming--macroscopic update of Section~\ref{sec:coupling}, which uses $(\rho^n,\bm u^n)$ together with $(\bm\sigma_p^{n+1},\bm F_p^{n+1})$ in collision before reconstructing $(\rho^{n+1},\bm u^{n+1},\nabla\bm u^{n+1})$ (with $\bm F_b+\bm F_p^{n+1}$ in the FD-Guo half-force shift). Unlike Eq.~\eqref{eq:shifted-proj}, the one-sided eigenvalue-floor control in Table~\ref{tab:clip-timeline} replaces only $a_1$ by $\max(a_1,\varepsilon)$ while leaving $a_2$ untouched, failing to restore positive definiteness whenever $a_1\leq a_2<0$.

For the smooth spectral-restoring source $\bm S_{\mathrm{cone}}(\A)$ in Eq.~\eqref{eq:Scone}, numerical evaluation prevents floating-point overflow in the softplus argument $z_k=-\gamma-a_k$ via the piecewise overflow guard
\[
s_k=\begin{cases}
0, & z_k\leq -40\delta_{\mathrm{cone}},\\
\dfrac{\delta_{\mathrm{cone}}}{\tau_{\mathrm{cone}}}\ln\!\big(1+e^{z_k/\delta_{\mathrm{cone}}}\big), & -40\delta_{\mathrm{cone}}<z_k\leq 40\delta_{\mathrm{cone}},\\
\dfrac{z_k}{\tau_{\mathrm{cone}}}, & z_k>40\delta_{\mathrm{cone}},
\end{cases}
\]
which has $O(e^{-40}\delta_{\mathrm{cone}}/\tau_{\mathrm{cone}})$ splice offsets from the analytic $C^\infty$ kernel $g_{\mathrm{cone}}(a_k)$ at $u_k=z_k/\delta_{\mathrm{cone}}=\pm 40$, with $u_k\in(-40,40]$ evaluated in IEEE binary64 via the positive-argument overflow-safe identity $\ln(1+e^{-|u_k|})+\max(u_k,0)$ (truncating to $0$ once $1+e^{-|u_k|}$ rounds to $1$ below $u_k\approx -36.7$, within $O(10^{-16}\delta_{\mathrm{cone}}/\tau_{\mathrm{cone}})$ of Eq.~\eqref{eq:Scone}). Adding $\sum_{k=1}^2 s_k\bm q_k\bm q_k^{\!\top}$ to the constitutive stretching--relaxation tensor defines the total conformation source $S_a$, which enters conformation collision through $h_{ai}=w_i[1+(1-\Omega_{p1}/2)(\bm c_i\cdot\bm u)/c_s^2]S_a$ (with first-moment conformation TRT collision rate $\Omega_{p1}=1/\tau_p$) using the second-order Adams--Bashforth stencil $\Delta t(\frac{3}{2}h_{ai}^n-\frac{1}{2}h_{ai}^{n-1})$ initialized with $h_{ai}^{-1}=0$ (giving a first-step factor $\frac{3}{2}\Delta t\,h_{ai}^0$ when $h_{ai}^0\neq 0$, whereas $s_k^0=0$ for equilibrium starts $\A^0=\bm I$).

The auxiliary order-$8$ conformation filter in Table~\ref{tab:matrix} applies the periodic operator
\begin{equation*}
A_a\leftarrow A_a-\sigma_f(\mathcal D_x^4+\mathcal D_y^4)A_a,
\qquad
(\mathcal D_xq)_{mn}=\frac{2q_{mn}-q_{m-1,n}-q_{m+1,n}}{4},
\end{equation*}
(and analogously in $y$) with $\sigma_f=0.2$ at every step, writing increments back via Eq.~\eqref{eq:pop-writeback}. In the FD-TC polymer-force gain scan of Table~\ref{tab:matrix}, $\bm F_p$ in Eq.~\eqref{eq:forcing-kernels} is replaced by $g\bm F_p$ for $g\in\{0.50,0.70,0.85,1.00,1.15\}$ without altering Eq.~\eqref{eq:conf}.

\subsection{Sub-step stage diagnostics at first exit from the SPD cone}
\label{app:first-cross}

To bracket where discrete constraint relaxation first occurs within a single step of unmodified SM-1 ($\Wi=20$, $N=257$, $\mathrm{Sc}=10^5$, $\beta=2/3$), non-intrusive diagnostics evaluate $\det\A$ and $\lmin(\A)$ across the five stages of Table~\ref{tab:first-cross} without modifying solver arrays. Both minima first turn negative at step $35328$ ($t^*=4.99$) between collision and streaming, bracketing initial exit from $\mathcal K_h$ without isolating streaming alone since TRT collision already couples stretching, constitutive relaxation, and transport.

\begin{table}[!htbp]
\centering
\small
\begin{tabular}{@{}lrr@{}}
\toprule
Sub-step stage & $\min_{\bm x}\det\A$ & $\min_{\bm x}\lmin(\A)$ \\
\midrule
Step start                       & $ 2.52\times10^{-2}$ & $ 2.17\times10^{-5}$ \\
After collision                  & $ 3.34\times10^{-2}$ & $ 2.87\times10^{-5}$ \\
After streaming                  & $-4.31\times10^{-3}$ & $-3.71\times10^{-6}$ \\
After macroscopic reconstruction & $-4.31\times10^{-3}$ & $-3.71\times10^{-6}$ \\
Full-step end                    & $-4.31\times10^{-3}$ & $-3.71\times10^{-6}$ \\
\bottomrule
\end{tabular}
\caption{Five-stage sub-step diagnostic values at the first exit from $\mathcal K_h$ (step $35328$, $t^*=4.99$) in unmodified SM-1 ($\Wi=20$, $N=257$, $\mathrm{Sc}=10^5$, $\beta=2/3$). Because no projection is active and the hydrodynamic update does not overwrite $\A$, the final three stages coincide; minimizing nodes remain at the symmetry-paired points $(129,2)$ and $(129,256)$, whose post-streaming $\det\A$ and $\lmin(\A)$ values agree to within $3.1\times 10^{-15}$.}
\label{tab:first-cross}
\end{table}
\FloatBarrier

\subsection{Numerical verification and cross-build reproducibility}
\label{app:revision-checks}

Enabling per-step monitoring at $\beta=2/3$ leaves scheduled diagnostics unchanged and reproduces byte-identical $t^*\simeq 8$ snapshots. Across the shifted-cone and restoring series (Linux GNU Fortran $13.2.0$, \texttt{-O3 -fopenmp}), per-step checks inspect all $257^2$ nodal tensors over steps $0$--$850150$ across the twelve runs in Table~\ref{tab:shifted-cone} and their $\beta=2/3$ SM-1 counterparts, and to step $432185$ on the two $c_\gamma=0.5$ FD-TC runs (smooth restoring crosses $U_{\max}>50$ at step $429536$, $t^*=60.6297$). Near $0$ or $-\beta/(1-\beta)$, quadruple-precision (\texttt{real128}) checks verify exact $\lmin$ signs. All eleven $t^*=120$ hard-floor runs (six in Table~\ref{tab:shifted-cone}, one in Table~\ref{tab:matrix}, and the four $17$-digit checks below) satisfy $\lmin^{\mathrm{post}}\geq\ell_{\mathrm{floor}}$ within $10^{-12}\max(1,|\ell_{\mathrm{floor}}|)$ ($0$ violations), whereas the terminating FD-TC runs reach $9.54\times 10^{-9}$ at $c_\gamma=0$ ($4.6\times 10^{-10}$ below $\varepsilon$) and $-1+9.42\times 10^{-9}$ at $c_\gamma=0.5$ ($5.8\times 10^{-10}$ below $-1+\varepsilon$) from finite-precision $2\times 2$ spectral assembly conditioning before velocity-threshold crossing.

Rerunning both unmodified baseline-forcing $\beta=1/9$ cases on this Linux build with their archived $17$-digit inputs (which match the $12$-digit $\beta=2/3$ reference and differ by $|\Delta s_F|/s_F=5\times 10^{-13}$ and $|\Delta\mathrm{Sc}|/\mathrm{Sc}=2\times 10^{-9}$ from the nominal inputs of Table~\ref{tab:shifted-cone}) reproduces identical initial crossing steps ($20083$, $21077$ and $19690$, $21649$), crosses $U_{\max}>50$ at $t^*=45.98$ and $49.35$ (versus $44.94$ and $49.93$ in Table~\ref{tab:lowbeta-events}), and exits at $46.00$ and $50.00$, reflecting cross-build trajectory sensitivity at matched $17$-digit inputs rather than input rounding alone. On this binary (with thread-independent static loops), all six $17$-digit intervention runs ($c_\gamma\in\{0,0.5\}$ floors and $(c_\gamma,r_{\mathrm{cone}})=(0.5,0.01)$ restoring) reach $t^*=120$ ($0$ smooth clips; floor clips within $0.1\%$--$3.3\%$ of Table~\ref{tab:shifted-cone}; $t^*\in[60,120]$ mean $\Wi_x=49.4$ versus $66.3$ at force-matched $c_\gamma=0$).

\bibliographystyle{elsarticle-num}

\begin{thebibliography}{99}
\bibitem{DM1986}
F. Dupret, J. M. Marchal,
Loss of evolution in the flow of viscoelastic fluids,
J. Non-Newton. Fluid Mech. 20 (1986) 143--171.
\url{https://doi.org/10.1016/0377-0257(86)80019-2}.

\bibitem{Hulsen1990}
M. A. Hulsen,
A sufficient condition for a positive definite configuration tensor in differential models,
J. Non-Newton. Fluid Mech. 38 (1990) 93--100.
\url{https://doi.org/10.1016/0377-0257(90)85034-V}.

\bibitem{CK2012}
P. Constantin, M. Kliegl,
Note on global regularity for two-dimensional Oldroyd--B fluids with diffusive stress,
Arch. Ration. Mech. Anal. 206 (2012) 725--740.
\url{https://doi.org/10.1007/s00205-012-0537-0}.

\bibitem{Rutkevich}
I. M. Rutkevich,
Some general properties of the equations of viscoelastic incompressible fluid dynamics,
J. Appl. Math. Mech. 33 (1969) 30--39.
\url{https://doi.org/10.1016/0021-8928(69)90111-7}.

\bibitem{Rutkevich1970}
I. M. Rutkevich,
The propagation of small perturbations in a viscoelastic fluid,
J. Appl. Math. Mech. 34 (1970) 35--50.
\url{https://doi.org/10.1016/0021-8928(70)90006-7}.

\bibitem{JRS1985}
D. D. Joseph, M. Renardy, J.-C. Saut,
Hyperbolicity and change of type in the flow of viscoelastic fluids,
Arch. Ration. Mech. Anal. 87 (1985) 213--251.
\url{https://doi.org/10.1007/BF00250725}.

\bibitem{JS1990}
D. D. Joseph, J.-C. Saut,
Short-wave instabilities and ill-posed initial-value problems,
Theor. Comput. Fluid Dyn. 1 (1990) 191--227.
\url{https://doi.org/10.1007/BF00418002}.

\bibitem{GS1990}
C. Guillop\'e, J.-C. Saut,
Existence results for the flow of viscoelastic fluids with a differential constitutive law,
Nonlinear Anal. 15 (1990) 849--869.
\url{https://doi.org/10.1016/0362-546X(90)90097-Z}.

\bibitem{JS1986}
D. D. Joseph, J.-C. Saut,
Change of type and loss of evolution in the flow of viscoelastic fluids,
J. Non-Newton. Fluid Mech. 20 (1986) 117--141.
\url{https://doi.org/10.1016/0377-0257(86)80018-0}.

\bibitem{VC2003}
T. Vaithianathan, L. R. Collins,
Numerical approach to simulating turbulent flow of a viscoelastic polymer solution,
J. Comput. Phys. 187 (2003) 1--21.
\url{https://doi.org/10.1016/S0021-9991(03)00028-7}.

\bibitem{FK2004}
R. Fattal, R. Kupferman,
Constitutive laws for the matrix-logarithm of the conformation tensor,
J. Non-Newton. Fluid Mech. 123 (2004) 281--285.
\url{https://doi.org/10.1016/j.jnnfm.2004.08.008}.

\bibitem{FK2005}
R. Fattal, R. Kupferman,
Time-dependent simulation of viscoelastic flows at high Weissenberg number using the log-conformation representation,
J. Non-Newton. Fluid Mech. 126 (2005) 23--37.
\url{https://doi.org/10.1016/j.jnnfm.2004.12.003}.

\bibitem{Balci2011}
N. Balci, B. Thomases, M. Renardy, C. R. Doering,
Symmetric factorization of the conformation tensor in viscoelastic fluid models,
J. Non-Newton. Fluid Mech. 166 (2011) 546--553.
\url{https://doi.org/10.1016/j.jnnfm.2011.02.008}.

\bibitem{HL2007}
D. Hu, T. Leli\`evre,
New entropy estimates for Oldroyd--B and related models,
Commun. Math. Sci. 5 (2007) 909--916.
\url{https://doi.org/10.4310/CMS.2007.v5.n4.a9}.

\bibitem{HSA2021}
M. A. Hulsen, M. M. A. Spanjaards, P. D. Anderson,
Numerical stability of four positive \mbox{(semi-)}def\-i\-nite reformulations for viscoelastic fluid models in benchmark flows,
J. Non-Newton. Fluid Mech. 297 (2021) 104666.
\url{https://doi.org/10.1016/j.jnnfm.2021.104666}.

\bibitem{NoNeed}
D. Capocci, M. Linkmann, A. Morozov,
No need to stay positive: a practical approach to direct numerical simulations of elastic turbulence,
arXiv:2606.09468 (2026).
\url{https://arxiv.org/abs/2606.09468}.

\bibitem{OP2002}
R. G. Owens, T. N. Phillips,
Computational Rheology, Imperial College Press, London, 2002.

\bibitem{Chen2014}
X. Chen,
Numerical modeling of fluid--structure interaction with rheologically complex fluids,
Doctoral dissertation, Technische Universit\"at Darmstadt, 2014.
\url{https://doi.org/10.26083/tuprints-00004216}.

\bibitem{YuRapids2026}
Y. Yu,
Static compliance and directional instability in indefinite conformation states,
arXiv:2608.00032 (2026).
\url{https://arxiv.org/abs/2608.00032}.

\bibitem{Guo2002}
Z. Guo, C. Zheng, B. Shi,
Discrete lattice effects on the forcing term in the lattice Boltzmann method,
Phys. Rev. E 65 (2002) 046308.
\url{https://doi.org/10.1103/PhysRevE.65.046308}.

\bibitem{ZXW2025}
Y. Zhang, Q. Xu, B. Wen,
Time consistency of lattice Boltzmann equation with an external force,
Phys. Rev. E 112 (2025) 035306.
\url{https://doi.org/10.1103/7hkk-4kx9}.

\bibitem{Yu2026}
Y. Yu, S. Chen, Y. Zhou, L. Wang, H.-Z. Yuan, S. Shu,
An improved lattice Boltzmann method with a novel conservative boundary scheme for viscoelastic fluid flows,
J. Comput. Phys. 550 (2026) 114667.
\url{https://doi.org/10.1016/j.jcp.2026.114667}.

\bibitem{Oldroyd1950}
J. G. Oldroyd,
On the formulation of rheological equations of state,
Proc. R. Soc. Lond. A 200 (1950) 523--541.
\url{https://doi.org/10.1098/rspa.1950.0035}.

\bibitem{SB1995}
R. Sureshkumar, A. N. Beris,
Effect of artificial stress diffusivity on the stability of numerical calculations and the flow dynamics of time-dependent viscoelastic flows,
J. Non-Newton. Fluid Mech. 60 (1995) 53--80.
\url{https://doi.org/10.1016/0377-0257(95)01377-8}.

\bibitem{Ginzburg2008}
I. Ginzburg, F. Verhaeghe, D. d'Humi\`eres,
Two-relaxation-time lattice Boltzmann scheme: about parametrization, velocity, pressure and mixed boundary conditions,
Commun. Comput. Phys. 3 (2008) 427--478.
\url{https://doi.org/10.4208/cicp.2008.v3.p427}.

\bibitem{SYC2006}
X. Shan, X.-F. Yuan, H. Chen,
Kinetic theory representation of hydrodynamics: a way beyond the Navier--Stokes equation,
J. Fluid Mech. 550 (2006) 413--441.
\url{https://doi.org/10.1017/S0022112005008153}.

\bibitem{TS2007}
B. Thomases, M. Shelley,
Emergence of singular structures in Oldroyd--B fluids,
Phys. Fluids 19 (2007) 103103.
\url{https://doi.org/10.1063/1.2783426}.

\bibitem{BB2010}
S. Berti, G. Boffetta,
Elastic waves and transition to elastic turbulence in a two-dimensional viscoelastic Kolmogorov flow,
Phys. Rev. E 82 (2010) 036314.
\url{https://doi.org/10.1103/PhysRevE.82.036314}.

\bibitem{YuDataset2026}
[dataset] Y. Yu, L. Lian, F. Chu,
Data and code for ``Loss of positive definiteness as numerical relaxation in
high-Weissenberg-number computations'',
Zenodo, version 1.0.0 (2026).
\url{https://doi.org/10.5281/zenodo.21442104}.

\bibitem{GV2019}
A. Gupta, D. Vincenzi,
Effect of polymer-stress diffusion in the numerical simulation of elastic turbulence,
J. Fluid Mech. 870 (2019) 405--418.
\url{https://doi.org/10.1017/jfm.2019.224}.

\end{thebibliography}

\end{document}